\PassOptionsToPackage{sort}{natbib}
\documentclass[preprint,11pt,authoryear]{elsarticle}
\usepackage{lipsum}
\makeatletter
\def\ps@pprintTitle{%
	\let\@oddhead\@empty
	\let\@evenhead\@empty 
	\def\@oddfoot{}%
	\let\@evenfoot\@oddfoot}
\makeatother
\usepackage{amsmath,amsfonts,amsthm,amssymb} 
\DeclareMathOperator{\EX}{\mathbb{E}}  
\usepackage[]{algorithm} 
\usepackage[noend]{algpseudocode}
\usepackage{color}
\usepackage[table,xcdraw]{xcolor}
\usepackage{url}
\usepackage{graphicx}
\usepackage{fancyhdr}
\usepackage[margin=2.54cm]{geometry}

\usepackage{cases}
\usepackage{mathptmx} 
\usepackage[scaled=0.90]{helvet} 
\usepackage{courier} 
\normalfont
\usepackage[T1]{fontenc}

\usepackage{setspace} 
\algnewcommand\AND{\textbf{and }}
\algnewcommand\OR{\textbf{or }}

\definecolor{darkseagreen}{rgb}{0.56, 0.74, 0.56}
\definecolor{auburn}{rgb}{0.43, 0.21, 0.1}
\definecolor{applegreen}{rgb}{0.55, 0.71, 0.0}

\definecolor{green}{HTML}{00693E}

\usepackage{graphicx}
\usepackage{amssymb}

\begin{document}
	
	\begin{frontmatter}

		\title{Unmanned Aerial Vehicle Path Planning for Traffic Estimation and Detection of Non-Recurrent Congestion}

		\cortext[cor1]{Corresponding author\newline
			E-mail addresses: cesaryahia@utexas.edu (C.N. Yahia),  	shannonescott@utexas.edu (S.E. Scott), sboyles@austin.utexas.edu (S.D. Boyles),  	christian.claudel@utexas.edu (C.G. Claudel)  }
		
		\author[label1]{Cesar N. Yahia\corref{cor1}}
		\author[label2]{Shannon E. Scott}
		\author[label1]{Stephen D. Boyles}
		\author[label1]{and Christian G. Claudel}
		\address[label1]{Department of Civil, Architectural and Environmental Engineering, The University of Texas at Austin, 1 University Station C1761, Austin, TX 78712, United States}
		\address[label2]{Department of Aerospace Engineering and Engineering Mechanics, The University of Texas at Austin, 1 University Station C1761, Austin, TX 78712, United States \vspace{-25pt}}

		\begin{abstract}
			Unmanned aerial vehicles (UAVs) provide a novel means of extracting road and traffic information from video data. In particular, by analyzing objects in a video frame, UAVs can detect traffic characteristics and road incidents. Leveraging the mobility and detection capabilities of UAVs, we investigate a navigation algorithm that seeks to maximize information on the road/traffic state under non-recurrent congestion. We propose an active exploration framework that (1) assimilates UAV observations with speed-density sensor data, (2) quantifies uncertainty on the road/traffic state, and (3) adaptively navigates the UAV to minimize this uncertainty. The navigation algorithm uses the A-optimal information measure (mean uncertainty), and it depends on covariance matrices generated by a dual state ensemble Kalman filter (EnKF). In the EnKF procedure, since observations are a nonlinear function of the incident state variables, we use diagnostic variables that represent model predicted measurements. We also present a state update procedure that maintains a monotonic relationship between incident parameters and measurements. We compare the traffic/incident state estimates resulting from the UAV navigation-estimation procedure against corresponding estimates that do not use targeted UAV observations. Our results indicate that UAVs aid in detection of incidents under congested conditions where speed-density data are not informative.
		\end{abstract}
		
		
	\end{frontmatter}

	\section{Introduction}
	Non-recurrent congestion is caused by capacity-reducing incidents such as accidents, adverse weather conditions, and work zones. This type of congestion is considered to be the primary source of travel time variability and accounts for up to 30\% of congestion delay during peak periods \citep{skabardonis2003, anbaroglu2014, sun2019}. Thus, to minimize the impact of non-recurrent congestion, we need to rapidly detect incidents and allocate traffic management resources. \newline
	
	Recently, researchers demonstrated the use of unmanned aerial vehicles (UAVs) for traffic and incident monitoring \citep{krajewski2018, jin2016, stevens2017, lee2015, barmpounakis2019}. \cite{krajewski2018} extract vehicle trajectories from UAV video data. \cite{jin2016} use UAVs to map an incident site and determine the incident impact. Given the mobility and detection capabilities of UAVs, we develop an automatic path planning procedure that navigates UAVs towards informative traffic or incident observations.  \newline
	
	Traditional data-driven incident detection methods compare expected traffic conditions with sensor measurements. These algorithms detect that an incident occurred once collected data significantly deviates from expected conditions \citep{stephanedes1993}. However, such outlier-based methods suffer from random traffic fluctuations that cause false alarms. In addition, using data-driven methods, it is difficult to distinguish incident data from similar traffic patterns that occur due to congestion shock waves \citep{stephanedes1993, cheu1995, hawas2017}. \newline
	
	To improve incident detection, researchers explored estimation methods that use a macroscopic traffic model to jointly estimate traffic states and the incident severity. In particular, incident information can be integrated into model-based traffic estimation methods by modifying certain parameters (e.g. free flow speed and/or critical density) that reflect the incident impact \citep{wang2016, dabiri2015, wang2005b}. Alternatively, recent model-based estimation techniques rely on comparing the predictions of multiple traffic models with observed data; then, the estimation procedure seeks to identify the most likely model among a set of possible parameter configurations that represent different levels of incident severity  \citep{willsky1980,wang2014,wang2016,wang2016a}. These methods are promising but they are still limited in certain situations with poor observability where it is difficult to determine if speed-density measurements correspond to congestion under normal operating conditions or an actual reduction in road capacity.   \newline

\begin{figure}[H]
		\begin{center}
			\includegraphics[width=0.56\textwidth]{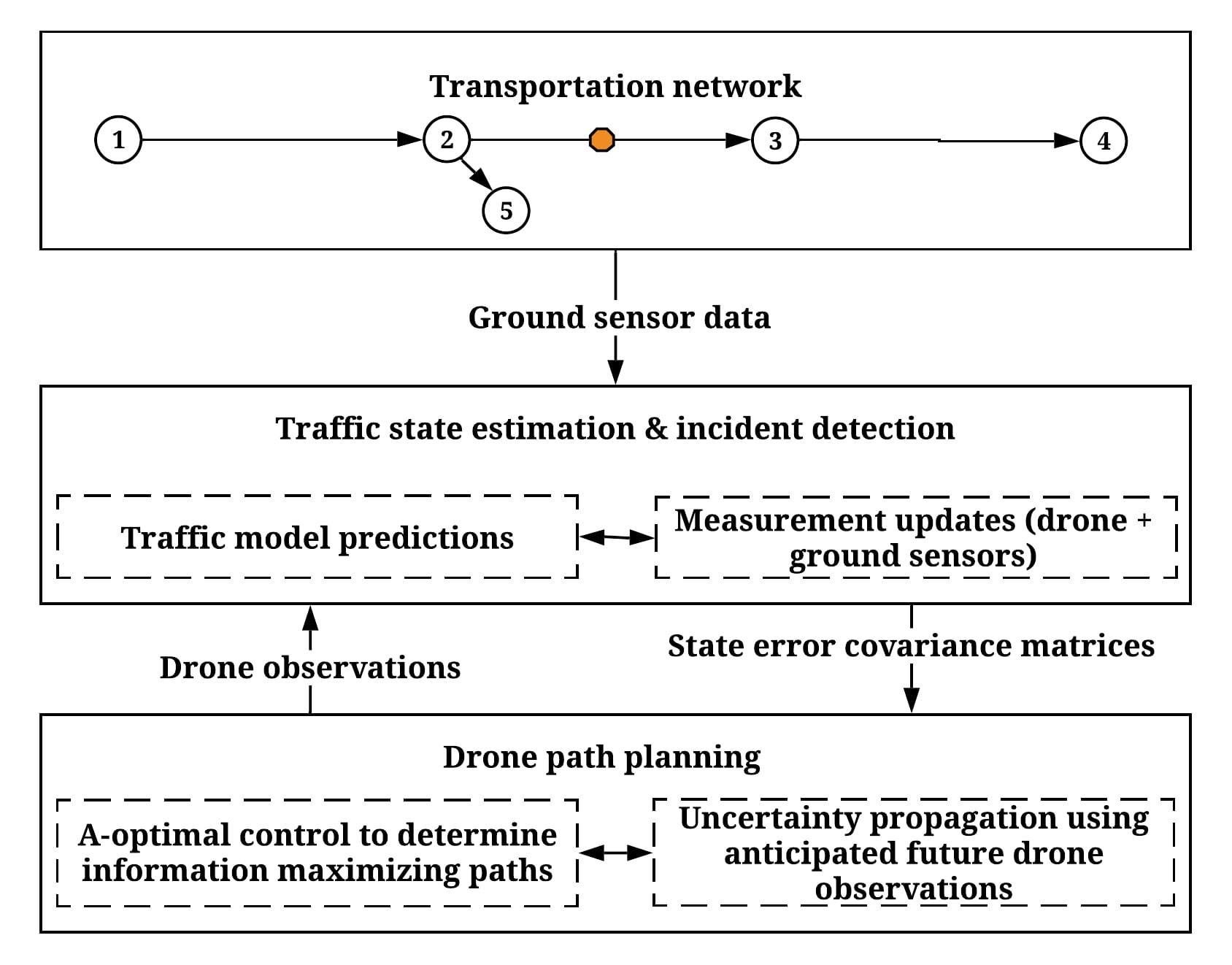}
		\end{center}
		\caption{Estimation and routing framework to navigate a UAV towards informative traffic state and incident observations.}
		\label{fig:fw}
	\end{figure}

	The objective of this article is to develop an estimation-planning framework that navigates a UAV towards informative observations of the underlying road/traffic conditions. The proposed framework (1) assimilates UAV density and capacity drop observations with local speed-density sensor measurements (2) quantifies the uncertainty on road and traffic states, and (3) adaptively navigates the UAV to minimize the uncertainty on state estimates. In particular, we develop an online one-step lookahead path planning algorithm that evaluates candidate UAV trajectories based on anticipated reduction of the \textit{mean uncertainty} (i.e., A-optimal path planning). The uncertainty is represented by time-varying covariance matrices that are generated by a dual ensemble Kalman filter (EnKF); these covariance matrices correspond to traffic densities and incident parameters (free flow speed and critical density). A key feature of the proposed estimation procedure is that we maintain a monotonic relationship between incident state parameters and observations. The proposed framework is shown in Figure \ref{fig:fw}.

	The remainder of this article is organized as follows. Section \ref{sec:lit} discusses the literature relevant to incident detection and traffic state estimation. In Section \ref{sec:estim}, we present the dual ensemble Kalman filter (EnKF) algorithm for simultaneous estimation of traffic densities and parameters. In Section \ref{sec:drone}, we discuss the difficulty in estimating capacity drops under congested conditions from traditional speed-density measurements, we quantify the uncertainty on the dual EnKF estimates, and we develop the framework shown in Figure \ref{fig:fw} to navigate a UAV towards uncertainty minimizing observations. In Section \ref{sec:results}, we present results obtained for a simulated freeway segment that show the advantage of generating targeted observations in congested conditions. Section \ref{sec:conc} concludes the article.
	
	\section{Literature Review}
	\label{sec:lit}
	The majority of incident detection methods rely on analyzing abnormalities in observed traffic data \citep{stephanedes1993}. These data-driven methods include threshold-based algorithms that have been applied since the 1970's. In particular, threshold-based algorithms compare patterns from detector observations to threshold values in a decision tree \citep{payne1976}. Alternative data-driven approaches such as time series analysis, artificial neural networks, and wavelet-based techniques were subsequently used to detect incidents \citep{cheu1995,teng2003,dia1997,stephanedes1995, stephanedes1993}. \cite{parkany2005} provide a comprehensive review on data-driven incident detection algorithms. The primary drawbacks of data-driven methods pertain to (1) fitting or specifying a large number of parameters, (2) difficulty in distinguishing incident traffic patterns from similar patterns that result from congestion under normal operating conditions, (3) susceptibility to random fluctuations in traffic data, and (4) difficulty in predicting the traffic state beyond locations where data is collected \citep{wang2016a, stephanedes1993, parkany2005}.\newline
	
	To detect incidents and simultaneously predict their impact on traffic conditions, researchers explored model-based estimation methods where model parameters reflect the incident severity. Incorporating the incident state in traffic estimation improves both incident detection capabilities and the resulting traffic state estimates \citep{wang2016}. \cite{wang2005b} proposed an extended Kalman filter that uses a macroscopic traffic flow model to estimate traffic densities as well as the free flow speed and critical density. They implemented joint state estimation where parameters and boundary variables are added to the state space, and they considered that flow and mean speed measurements could be obtained. Recent articles on simultaneous estimation of traffic states and fundamental diagram parameters include the use of count and trajectory data in a single optimization framework \citep{sun2017}. \newline

	Alternative model-based estimation techniques include methods that aim to identify the most likely traffic model among a set of models; in these methods, each model represents a different configuration of parameters \citep{wang2016,wang2016a}. In the case of incident detection, each model parametrization reflects a certain level of incident severity. The first article to consider this approach used an extended Kalman filter to select the most likely model \citep{willsky1980}. This framework was then enhanced to allow for dependencies between the most likely models chosen across time \citep{wang2016}. In particular, given pre-specified incident evolution dynamics, an interactive multiple model ensemble Kalman filter and a multiple model particle filter were used to simultaneously estimate traffic states and the incident severity  \citep{wang2014, wang2016, wang2016a}. \newline
	
	In model-based estimation methods that use a macroscopic traffic model, for typical sensor data such as speed-density measurements, it is difficult to distinguish between traffic patterns that result from incidents and similar patterns observed under incident-free congestion. In particular, whether there is an incident or not, we will observe low speed and high density measurements under congested conditions. This poor observability is further discussed in Section \ref{sec:drone}.\newline

	Thus, we propose an estimation-planning framework that navigates a UAV towards informative traffic/incident state observations. First, we present a dual state EnKF estimation procedure that generates Gaussian distributions reflecting the uncertainty on traffic/parameter state estimates. Then, we navigate the UAV towards observations that minimize the mean uncertainty; effectively, the UAV is navigated towards congested incident locations where it is difficult to infer the incident state from speed-density data. We note that the proposed path planning procedure can be used to navigate a mobile sensor towards uncertainty minimizing observations without updating parameters that represent incident severity. Similarly, the proposed estimation procedure (that maintains a monotonic relationship between measurements and incident parameters) can be used without the additional uncertainty-minimizing UAV observations.

	\section{Traffic State \& Parameter Estimation under Non-Recurrent Congestion}
	\label{sec:estim}
	In this section, we implement a dual state ensemble Kalman filter. The dual EnKF will result in separate traffic and parameter state covariance matrices. These matrices represent the uncertainty on the traffic/parameter state estimates. In Section \ref{sec:drone}, the uncertainty will be quantified along candidate UAV paths to determine the UAV trajectory that maximizes traffic/incident information.\newline
	
	A key feature of the proposed estimation procedure is that it maintains a monotonic relationship between observations and parameter state variables; advantages of this monotonicity are further discussed in Section \ref{sec:params}. In addition, since the measurements are not linearly related to the parameters of interest, we use model predicted measurements that represent anticipated observations for the given parameter values. \newline
	
	The non-recurrent congestion incidents we consider do not significantly impact the jam density. This type of incidents could represent adverse weather conditions, roadside accidents, or work zones. Compared to existing methods that consider lane-blocking incidents or a fixed incident free flow speed \citep{wang2014, wang2016, wang2016a, lu2013}, we aim to analyze the variation in free flow speed and critical density.

	\subsection{The Dual State Ensemble Kalman Filter}
	The dual EnKF is composed of two separate ensemble Kalman filters for traffic densities and parameters (free flow speed and critical density) working in parallel. Each EnKF is a stochastic filter that propagates ensemble members (samples) representing the state statistics \citep{evensen2003, evensen2009,blandin2012}. The filters interact by recursively feeding best estimates into each other at every update step. In particular, the updated parameters are used to adjust the forward model of the traffic densities EnKF; similarly, the resulting traffic state estimates inform subsequent parameter updates. In each filter, the ensemble mean is the best estimate on the true traffic/parameter state and the ensemble covariance corresponds to the error on the ensemble mean \citep{evensen2003, evensen2009}.\newline
	
	As a model-based estimation technique, the dual EnKF is limited by the accuracy of the traffic model in reflecting driver behavior and traffic dynamics. We use a triangular flow-density relationship that disregards driver heterogeneity and car-following behavior. This simple representation of traffic dynamics enables efficient prediction of traffic densities in the EnKF procedure.   \newline
	
	The traffic state is represented by densities $\rho(x,t)$ propagated forward using the cell transmission model. The incident severity is represented by the free flow speed $u_{f}$ and critical density $\rho_{cr}$ parameters at incident prone locations. The $u_{f}$ parameters are propagated forward using a random walk. On the other hand, the $\rho_{cr}$ parameters are updated based on corresponding $u_{f}$ updates; this parameter update procedure aims to maintain a monotonic relationship between parameters ($u_{f}$ and $\rho_{cr}$) and speed-density observations. Equivalently, we can propagate $\rho_{cr}$ using a random walk and update $u_{f}$ based on the corresponding $\rho_{cr}$ updates. We consider that the traffic state is directly observed using loop detector density measurements. We also consider that, for the given best estimate on traffic densities, the incident parameters are observed using less frequent speed measurements. \newline

	While alternative optimization methods could be used to specify the estimation objectives \citep{canepa2017}, the dual EnKF is a variance-minimizing scheme that enables efficient updating of Gaussian covariance matrices. In addition, compared to methods that estimate the most likely traffic model \citep{willsky1980, wang2016, wang2016a}, the dual EnKF uses continuous variables for incident parameters; thus, the dual EnKF estimates are not limited to a predefined set of incident severity levels.\newline

	Importantly, a dual estimation procedure enables us to maintain separate covariance matrices for traffic states and parameter estimates. Maintaining separate covariance matrices is a critical component of the proposed UAV navigation algorithm (Section \ref{sec:drone}); precisely, the UAV navigation algorithm identifies targeted uncertainty minimizing measurements based on the \textit{relative} uncertainty between traffic and parameter state estimates.\newline
	
	\subsection{Traffic State EnKF}
	In the traffic densities EnKF, the Lighthill-Whitham-Richards partial differential equation (LWR PDE) is used to represent traffic dynamics. This PDE is shown in Equation \ref{eqn:lwr} where $\rho(x,t)$ and $v(\rho(x,t))$ are the density and velocity at a particular point in space and time, respectively. Following \cite{wang2014}, we use a speed-density relationship that corresponds to a triangular flow-density diagram (Equation \ref{eqn:model}). In this equation, $\rho_{cr}$ is the critical density, $\rho_{j}$ is the jam density, and $u_{f}$ is the free flow speed. For implementation, the LWR PDE is discretized using a Godunov scheme to obtain the cell transmission model (CTM) \citep{daganzo1994a, daganzo1995a, godunov1959}. Thus, as the forward model in the traffic densities EnKF, the CTM will be used to propagate traffic flow through the network (i.e., the CTM will be used to track densities $\rho(x,t)$ across time). 
	
	\begin{spacing}{1.25}
	\begin{equation}
	\frac{\partial \rho(x,t) }{\partial t} + \frac{\partial (\rho(x,t) v(\rho(x,t)) ) }{\partial x} =0
	\label{eqn:lwr}
	\end{equation}
	\begin{equation}
	v(\rho(x,t))=
	\begin{cases}
	u_{f} & \text{for } \rho(x,t) \leq \rho_{cr} \\
	\frac{u_{f}\rho_{cr}(\rho_{j}- \rho(x,t))}{ \rho(x,t)(\rho_{j}-\rho_{cr})} & \text{otherwise }
	\end{cases}
	\label{eqn:model}
	\end{equation}
	\end{spacing}

	The resulting traffic densities EnKF is shown in Equations \ref{eqn:firstTraffic}--\ref{eqn:lastTraffic} \citep{evensen2003}. The $M_{\rho}\times1$ vector $\mathbf{\rho}_{i}^{t}$ represents densities associated with ensemble member $i$ at time $t$, where $M_{\rho}$ is the number of density parameters to be estimated across the network. The forward model $\mathsf{CTM_{\Delta t}}$ propagates traffic densities from time $t$ until time $t+\Delta t$ using the CTM. The model errors $w$ is a vector of Gaussian white noise such that $w\sim\text{N}(\mathbf{0},\mathbf{Q_{\rho}})$. For $N$ ensemble members, $\mathbf{A}$ is an $M_{\rho}\times N$ matrix that stores the ensemble members in its columns, $\mathbf{1_{N}}$ is an $N\times N$ scale matrix such that every element is $1/N$, and  $\mathbf{\bar{A}}$ is an $M_{\rho}\times N$ matrix where every column is the ensemble mean. \newline
	
	In terms of observations, $\mathbf{d_{j}}$ is an $M_{\rho}\times 1$ vector representing a particular perturbation of the vector of density measurements $\mathbf{d}^{\rho}$ using Gaussian white noise observation errors $\epsilon^{\rho}\sim\text{N}(\mathbf{0},\mathbf{R_{\rho}})$. The model and observation errors are independent of each other. The $M_{\rho}\times N$ matrix $\mathbf{D}$ stores the perturbed observations. The observation errors are stored in the $M_{\rho}\times N$ matrix $\Upsilon$. The matrix $\mathbf{H}$ is an observation matrix; in the traffic densities EnKF, $\mathbf{H}$ is the $M_{\rho}\times M_{\rho}$ identity matrix since state variables (densities) are directly measured.\newline
	
	After every update stage, the updated ensemble members $\mathbf{\rho}_{i}^{t+\Delta t}$ are stored in the columns of the $M_{\rho}\times N$ matrix $\mathbf{A^{a}}$. The updated ensemble covariance matrix is $\mathbf{P}$. Thus, $\mathbf{P}$ represents the uncertainty on the traffic density estimates.

	\begin{spacing}{1.25}
		\begin{equation}
		\mathbf{\rho}_{i}^{t+\Delta t|t} = \mathsf{CTM_{\Delta t}}(\mathbf{\rho}_{i}^{t}) + w \quad \forall i \in \{1,..,N \}
		\label{eqn:firstTraffic}
		\end{equation}
		\begin{equation}
		\mathbf{A}=\left[\mathbf{\rho}_{1}^{t+\Delta t|t}, \mathbf{\rho}_{2}^{t+\Delta t|t}, ...,\mathbf{\rho}_{N}^{t+\Delta t|t}\right]
		\label{eqn:secTraffic}
		\end{equation}
		\begin{equation}
		\mathbf{\bar{A}}=\mathbf{A}\mathbf{1_{N}}
		\end{equation}
		\begin{equation}
		\mathbf{d_{j}} = \mathbf{d}^{\rho} + \epsilon_{j}^{\rho}  \quad \forall j \in \{1,..,N \}
		\label{eqn:denMes}
		\end{equation}
		\begin{equation}
		\mathbf{D}=\left[\mathbf{d_{1}}, \mathbf{d_{2}}, ...,\mathbf{d_{N}}\right]
		\end{equation}
		\begin{equation}
		\Upsilon=\left[\epsilon_{1}^{\rho}, \epsilon_{2}^{\rho}, ...,\epsilon_{N}^{\rho}\right]
		\end{equation}
		\begin{equation}
		\mathbf{A^{a}}=\mathbf{A}+(\mathbf{A}-\mathbf{\bar{A}})(\mathbf{A}-\mathbf{\bar{A}})^\top\mathbf{H}^\top(\mathbf{H}(\mathbf{A}-\mathbf{\bar{A}})(\mathbf{A}-\mathbf{\bar{A}})^\top\mathbf{H}^\top + \Upsilon \Upsilon^\top   )^{-1}(\mathbf{D}-\mathbf{H}\mathbf{A})
		\label{eqn:updateEqH}
		\end{equation}
		\begin{equation}
		\mathbf{P}= \frac{1}{N-1} (\mathbf{A^{a}}-\mathbf{A^{a}}\mathbf{1_{N}})(\mathbf{A^{a}}-\mathbf{A^{a}}\mathbf{1_{N}})^\top
		\label{eqn:lastTraffic}
		\end{equation}
	\end{spacing}

	\subsection{Parameters EnKF: Free Flow Speed and Critical Density Updates}
	\label{sec:params}
	In addition to the traffic density estimates, we present a parameters EnKF for propagating and updating parameter estimates once speed measurements are obtained. Given the best estimate on traffic densities, we can relate the observed speed measurements $v(\rho(x,t))$ to $u_{f}$ and $\rho_{cr}$ through Equation \ref{eqn:model}. Clearly, the observation function relating measurements to parameters is nonlinear; this nonlinearity implies that we can not construct an observation matrix $\mathbf{H}$ similar to the matrix that appears in Equation \ref{eqn:updateEqH}. Thus, we discuss an approach for incorporating the non-linear parameter-measurement relationship using model diagnostic variables; in particular, the diagnostic variables represent model predicted measurements. We also present a parameter update procedure where either $\rho_{cr}$ or $u_{f}$ is updated by the EnKF, and the other parameter is subsequently updated using a predefined relationship between $\rho_{cr}$ and $u_{f}$. This parameter update procedure preserves a monotonic relationship between parameter updates and observations. \\
	
	To incorporate the non-linear parameter-measurements observation function, we introduce the matrix $\mathbf{\hat{A}}$ \citep{evensen2009}. The columns of $\mathbf{\hat{A}}$ are predicted velocities $v(\rho(x,t))$ for the current $\rho_{cr}$ and $u_{f}$ values (Equation \ref{eqn:newA}); these predicted velocities are computed by plugging in $\rho_{cr}$ and $u_{f}$ in Equation \ref{eqn:model}. For ensemble members $i$ at time $t$, we use $\mathsf{M}(\mathbf{u}_{f, i}^{t}, \mathbf{\rho}_{cr, i}^{t})$ to denote this non-linear function relating the parameters to the predicted velocity. Thus,  $\mathsf{M}(\mathbf{u}_{f, i}^{t}, \mathbf{\rho}_{cr, i}^{t})$ is a vector of model predicted velocities at incident prone locations, where the predicted velocities correspond to the parameters ($\mathbf{\rho}_{cr, i}^{t}$ and $\mathbf{u}_{f, i}^{t}$) in ensemble member $i$ at time $t$. Note that $\mathbf{u}_{f, i}^{t}$ is an $M_{u_{f}}\times1$ vector of free flow speeds $u_{f}$ at incident prone locations ($M_{u_{f}}$ is the number of $u_{f}$ parameters to be estimated). Similarly, $\mathbf{\rho}_{cr, i}^{t}$ is a vector of critical densities at incident prone locations.
	
	\begin{equation}
	\mathbf{\hat{A}}=\left[\mathsf{M}(\mathbf{u}_{f, 1}^{t},\mathbf{\rho}_{cr, 1}^{t}), \mathsf{M}(\mathbf{u}_{f, 2}^{t},\mathbf{\rho}_{cr, 2}^{t}), ...,\mathsf{M}(\mathbf{u}_{f, N}^{t},\mathbf{\rho}_{cr, N}^{t})\right]
	\label{eqn:newA}
	\end{equation}
	
	Let $\left(\mathbf{p}_{i}^{t}\right)^\top=\left[ \left(\mathbf{u}_{f, i}^{t}\right)^\top, \left(\mathbf{\rho}_{cr, i}^{t} \right)^\top  \right]  $ be the vector of all parameters that are associated with ensemble member $i$ at time $t$. In addition, let $\mathbf{F}$ be a model that propagates the parameters forward in time ($\mathbf{F}$ includes model errors). Then, the resulting EnKF equations would be as shown in Equations \ref{eqn:firstParam}--\ref{eqn:lastParam}. The primary difference between the parameter EnKF equations and the traffic densities EnKF is the use of $\mathbf{\hat{A}}$ in the update Equation \ref{eqn:newUpdate}. In addition, $\mathbf{d_{j}}$ in Equation \ref{eqn:speedMes} is an $M_{v}\times1$ vector representing a particular perturbation of the of \textit{speed} measurements $\mathbf{d}^{v}$ using Gaussian observation errors $\epsilon^{v}\sim\text{N}(\mathbf{0},\mathbf{R_{v}})$ (the dimension of the measurement vector $M_{v}$ may be different from the number of estimated parameters).
	\begin{spacing}{1.25}
		\begin{equation}
		\mathbf{p}_{i}^{t+\Delta t|t} = \mathbf{F}(\mathbf{p}_{i}^{t}) \quad \forall i \in \{1,..,N \}
		\label{eqn:firstParam}
		\end{equation}
		\begin{equation}
		\mathbf{A}=\left[\mathbf{p}_{1}^{t+\Delta t|t}, \mathbf{p}_{2}^{t+\Delta t|t}, ...,\mathbf{p}_{N}^{t+\Delta t|t}\right]
		\end{equation}
		\begin{equation}
		\mathbf{\bar{A}}=\mathbf{A}\mathbf{1_{N}}
		\label{eqn:Abarparam}
		\end{equation}
		\begin{equation}
		\mathbf{d_{j}} = \mathbf{d}^{v} + \epsilon_{j}^{v}  \quad \forall j \in \{1,..,N \}
		\label{eqn:speedMes}
		\end{equation}
		\begin{equation}
		\mathbf{D}=\left[\mathbf{d_{1}}, \mathbf{d_{2}}, ...,\mathbf{d_{N}}\right]
		\end{equation}
		\begin{equation}
		\Upsilon=\left[\epsilon_{1}^{v}, \epsilon_{2}^{v}, ...,\epsilon_{N}^{v}\right]
		\end{equation}
		\begin{equation}
		\mathbf{A^{a}}=\mathbf{A}+(\mathbf{A}-\mathbf{\bar{A}})(\mathbf{\hat{A}}-\mathbf{\hat{A}}\mathbf{1_{N}})^\top((\mathbf{\hat{A}}-\mathbf{\hat{A}}\mathbf{1_{N}})(\mathbf{\hat{A}}-\mathbf{\hat{A}}\mathbf{1_{N}})^\top + \Upsilon \Upsilon^\top   )^{-1}(\mathbf{D}-\mathbf{\hat{A}})
		\label{eqn:newUpdate}
		\end{equation}
		\begin{equation}
		\mathbf{P}= \frac{1}{N-1} (\mathbf{A^{a}}-\mathbf{A^{a}}\mathbf{1_{N}})(\mathbf{A^{a}}-\mathbf{A^{a}}\mathbf{1_{N}})^\top
		\label{eqn:lastParam}
		\end{equation}
	\end{spacing}
	
\begin{figure}[H]
		\begin{center}
			\includegraphics[width=0.53\textwidth]{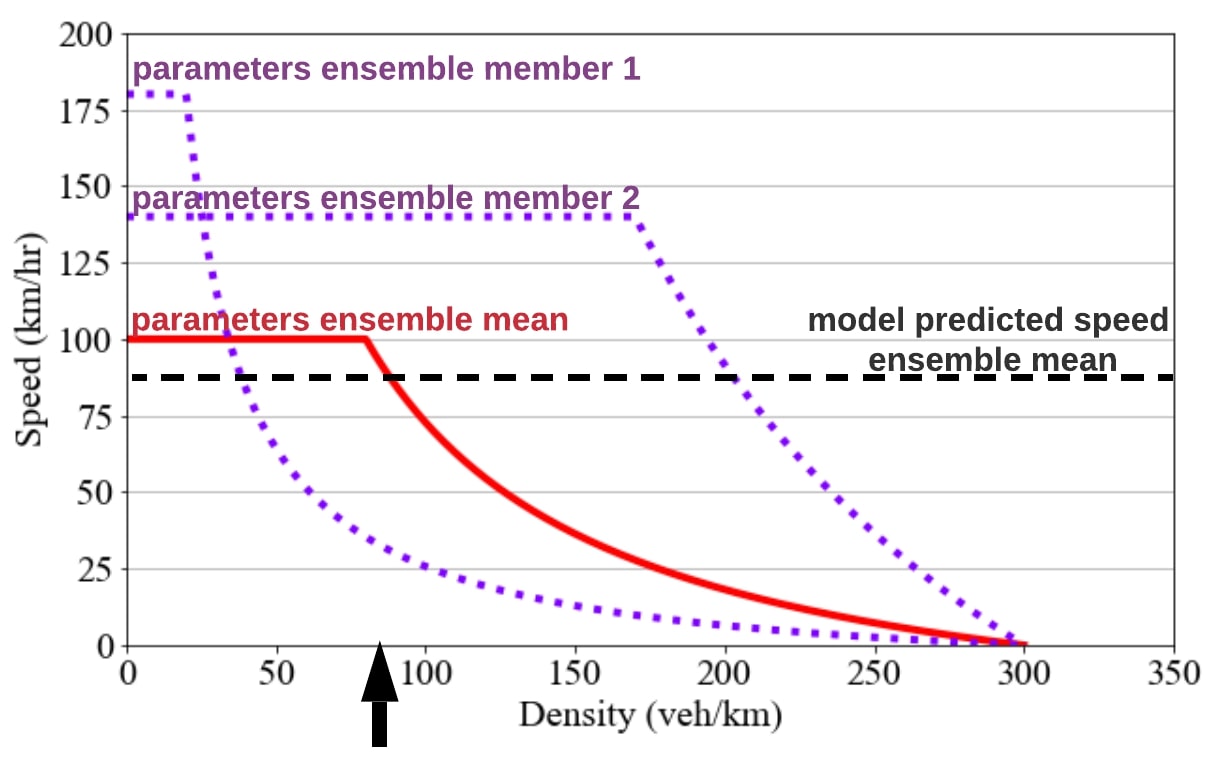}
		\end{center}
		\caption{Non-monotonic relationship between parameters and model predicted measurements. Dotted lines: speed-density relationship at a particular incident location for different ensemble members of the parameters EnKF (to facilitate illustration, we only exhibit two ensemble members). Solid line: speed-density relationship associated with the ensemble mean. Dashed line: ensemble mean of the model predicted speed measurements at the incident location.  }
		\label{fig:non-monoto}
	\end{figure}

	The modified update procedure in Equation \ref{eqn:newUpdate} works well when the $\mathsf{M}(\cdot)$ functions are monotonic and not highly nonlinear \citep{evensen2009}. If  $\mathsf{M}(\cdot)$ is non-monotonic, then it would not be clear if the EnKF should increase or decrease the parameter values in response to observed measurements. In more detail, Figure \ref{fig:non-monoto} illustrates the speed-density relationships (dotted lines) at a particular incident location. Each speed-density relationship corresponds to $u_{f}$ and $\rho_{cr}$ parameters of a specific ensemble member that is stored in $\mathbf{A}$. To facilitate illustration, we show parameter values representing only two ensemble members at one incident prone location. The solid line in Figure \ref{fig:non-monoto} shows the parameters ensemble mean at the incident location ($\rho_{cr}$ and $u_{f}$ elements of a column in $\mathbf{\bar{A}}$).\newline
	
	Consider that the density value assimilated through the traffic densities EnKF is $80$veh/km (as marked using an upward pointing arrow). At this density value, we can determine the model predicted speed associated with each ensemble speed-density relationship; these model predicted speeds will be stored in $\mathbf{\hat{A}}$ (the model predicted speed for ensemble member 1 is $35$km/hr and the corresponding value for ensemble member 2 is $140$km/hr). The mean of the model predicted speeds across ensemble members is shown using a dashed line (i.e., the incident location component of a column in $\mathbf{\hat{A}}\mathbf{1_{N}}$).\newline
	
	Observe that in Figure \ref{fig:non-monoto} the relationship between the parameter state variables and the model predicted speeds is \textit{non-monotonic}. In particular, an increase in $u_{f}$ above its ensemble mean value may result in low model predicted speed (ensemble member 1) or high model predicted speed (ensemble member 2). This non-monotonic relationship impacts the $(\mathbf{A}-\mathbf{\bar{A}})(\mathbf{\hat{A}}-\mathbf{\hat{A}}\mathbf{1_{N}})^\top$term in Equation \ref{eqn:newUpdate}, where $(\mathbf{A}-\mathbf{\bar{A}})(\mathbf{\hat{A}}-\mathbf{\hat{A}}\mathbf{1_{N}})^\top$ captures the covariance between parameter states and model predicted speeds. Specifically, it would not be clear if larger values of $u_{f}$ (relative to the $u_{f}$ ensemble mean) are associated with higher or lower model predicted speeds (relative to the model predicted speed ensemble mean). This implies that for a certain deviation of the model predicted speed from the observed measurements ($\mathbf{D}-\mathbf{\hat{A}}$), it would not be clear whether the EnKF should increase or decrease the $u_{f}$ state variables so that the model predicted speeds would match the observed measurements.\newline

\begin{figure}[H]
		\begin{center}
			\includegraphics[width=0.54\textwidth]{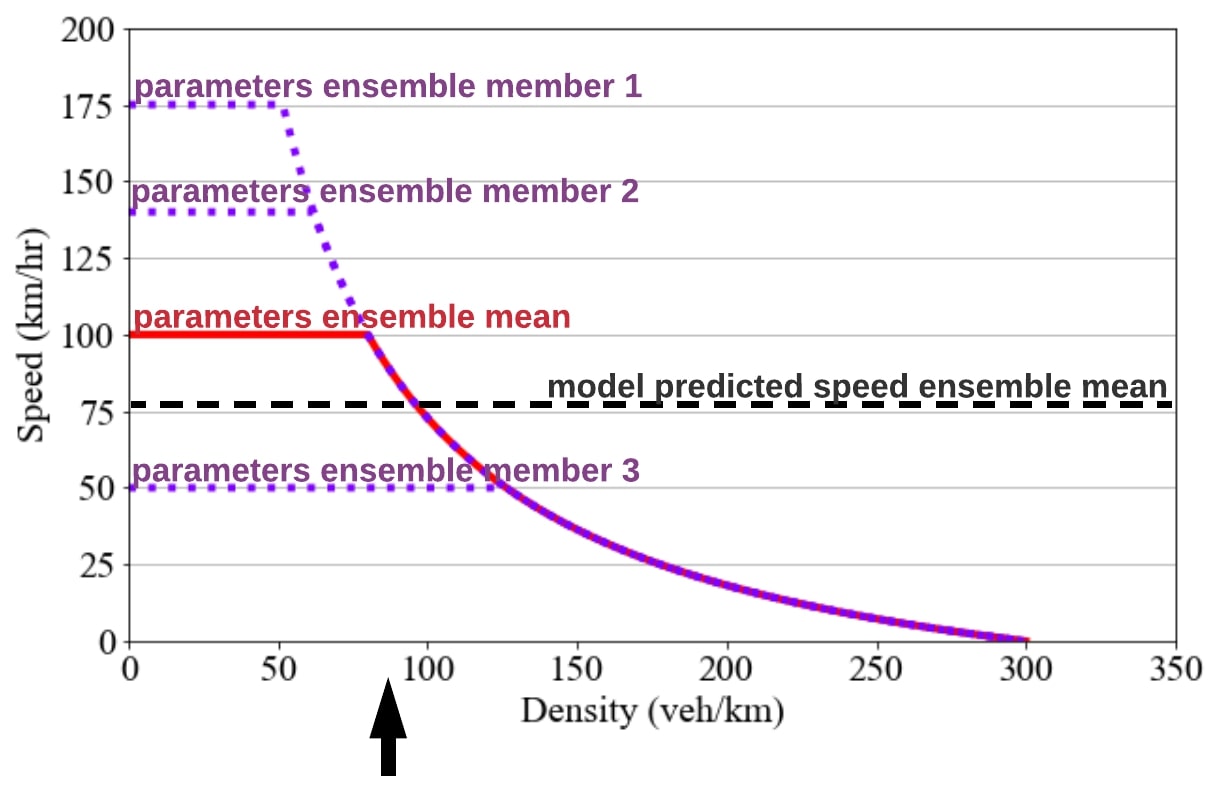}
		\end{center}
		\caption{Monotonic relationship between parameters and model predicted measurements. Dotted lines: speed-density relationship at a particular incident location for different ensemble members of the parameters EnKF (to facilitate illustration, we only exhibit three ensemble members). Solid line: speed-density relationship associated with the ensemble mean. Dashed line: ensemble mean of the model predicted speed measurements at the incident location.  }
		\label{fig:monotone}
	\end{figure}

	Alternatively, in Figure \ref{fig:monotone}, we enforce a relationship between $u_{f}$ and $\rho_{cr}$ across ensemble members. As shown, at the density value assimilated through the traffic densities EnKF ($80$veh/km), the model predicted speed is a non-decreasing function of $u_{f}$. This implies that in terms of the covariance $(\mathbf{A}-\mathbf{\bar{A}})(\mathbf{\hat{A}}-\mathbf{\hat{A}}\mathbf{1_{N}})^\top$, greater values of $u_{f}$ will be associated with greater values of model predicted speed. Similarly, we observe that the model predicted speed is a monotonic function of $\rho_{cr}$. In particular, the model predicted speed is a non-increasing function of $\rho_{cr}$. Thus, if the observed measurement is greater than the model predicted speed, the EnKF tends to increase $u_{f}$ and decrease $\rho_{cr}$ to match the measurement. On the other hand, if the observed measurement is less than the model predicted speed, the EnKF tends to decrease $u_{f}$ and increase $\rho_{cr}$ to match the measurement. However, for each ensemble member, $u_{f}$ is coupled with $\rho_{cr}$ through a specific relationship; this coupling between $u_{f}$ and $\rho_{cr}$ results in a monotonic relationship between parameters and observations.
		
	In more detail, to achieve the monotonic relationship between parameters and observations, we maintain a fixed backward wave speed $w^{0}$ across ensembles at each each incident prone location (as shown in Figure \ref{fig:monotone}). Let $\rho_{cr}^{0}$ and $u_{f}^{0}$ be the parameters associated with the backward wave speed $w^{0}$; these parameters and the corresponding backward wave speed $w^{0}$ may be calibrated during incident-free conditions from prior data. In each ensemble, for a specific incident prone location, the critical density $\rho_{cr}^{t}$ at any time $t$ should be related to the free flow speed $u_{f}^{t}$ via Equation \ref{eqn:ufrhoRel}. This equation can be further simplified into Equation \ref{eqn:ufrhoRelSimple}.

	\begin{spacing}{1.25}
	\begin{equation}
		\rho_{cr}^{t} = \frac{\rho_{cr}^{0} u_{f}^{0} \rho_{j}}{u_{f}^{t}(\rho_{j}-\rho_{cr}^{0}) + \rho_{cr}^{0}u_{f}^{0} }
		\label{eqn:ufrhoRel}
	\end{equation}
		
	\begin{equation}
		w^{0}=\frac{\rho_{cr}^{0} u_{f}^{0}}{\rho_{j}-\rho_{cr}^{0}}
		\label{eqn:w0}
	\end{equation}
	\begin{equation}
		\rho_{cr}^{t} = \frac{\rho_{j}w^{0}}{u_{f}^{t} + w^{0} }
		\label{eqn:ufrhoRelSimple}
	\end{equation}
	
	\end{spacing}
	
	Thus, using the relationship between $u_{f}^{t}$ and $\rho_{cr}^{t}$ in Equation \ref{eqn:ufrhoRelSimple}, we can express one parameter in terms of the other and substitute in Equation \ref{eqn:model}. Subsequently, we obtain a \textit{monotonic} non-linear observation function that relates the model predicted speed to either of the parameters. In particular, if we express $\rho_{cr}^{t}$ in terms of $u_{f}^{t}$ and substitute the resulting expression into Equation \ref{eqn:model}, we get Equation \ref{eqn:modeluf}. In turn, Equation \ref{eqn:modeluf} can be simplified into an $\mathsf{M}(\cdot)$ function that only depends on $u_{f}^{t}$ as shown in Equation \ref{eqn:Mspeed}. Notice that Equation \ref{eqn:Mspeed} monotonically relates predicted speeds $v(\rho(x, t))$ to $u_{f}^{t}$ such that an increase in $u_{f}^{t}$ is associated with non-decreasing model predicted speed. 
	
	\begin{spacing}{1.25}
	\begin{equation}
	v(\rho(x,t))=
	\begin{cases}
	u_{f}^{t} & \text{for } \rho(x,t) \leq \frac{\rho_{j}w^{0}}{u_{f}^{t} + w^{0} } \\
	\frac{u_{f}^{t}\left(\frac{\rho_{j}w^{0}}{u_{f}^{t} + w^{0} }\right)(\rho_{j}- \rho(x,t))}{ \rho(x,t)\left(\rho_{j}-\left(\frac{\rho_{j}w^{0}}{u_{f}^{t} + w^{0} }\right)\right)} & \text{otherwise }
	\end{cases}
	\label{eqn:modeluf}
	\end{equation}
	\begin{equation}
	\mathsf{M}(u_{f}^{t})=
	\begin{cases}
	u_{f}^{t} & \text{for } \rho(x,t) \leq \frac{\rho_{j}w^{0}}{u_{f}^{t} + w^{0} } \\
	w^{0}\left(\frac{\rho_{j}-\rho(x,t)}{\rho(x,t)}\right) & \text{otherwise }
	\end{cases}
	\label{eqn:Mspeed}
	\end{equation}
	\end{spacing}

	Equivalently, we can choose to express $u_{f}^{t}$ in terms of $\rho_{cr}^{t}$ (Equation \ref{eqn:parameterRel}) and substitute the resulting expression into Equation \ref{eqn:model}; this leads to Equation \ref{eqn:modelrhocr} that can be simplified into an $\mathsf{M}(\cdot)$ function that only depends on $\rho_{cr}^{t}$ as shown in Equation \ref{eqn:Mspeedrhocr}. This function also monotonically relates $\rho_{cr}^{t}$ to the model predicted speed such that an increase in $\rho_{cr}^{t}$ is associated with non-increasing model predicted speed.

	\begin{spacing}{1.25}
		\begin{equation}
		u_{f}^{t} = \frac{w^{0}\left(\rho_{j}-\rho_{cr}^{t}\right)}{\rho_{cr}^{t}}
		\label{eqn:parameterRel}
		\end{equation}
		\begin{equation}
		v(\rho(x,t))=
		\begin{cases}
		\frac{w^{0}\left(\rho_{j}-\rho_{cr}^{t}\right)}{\rho_{cr}^{t}} & \text{for } \rho(x,t) \leq \rho_{cr}^{t} \\
		\frac{\left( \frac{w^{0}\left(\rho_{j}-\rho_{cr}^{t}\right)}{\rho_{cr}^{t}} \right)\rho_{cr}^{t}(\rho_{j}- \rho(x,t))}{ \rho(x,t)(\rho_{j}-\rho_{cr}^{t})} & \text{otherwise }
		\end{cases}
		\label{eqn:modelrhocr}
		\end{equation}
		\begin{equation}
		\mathsf{M}(\rho_{cr}^{t})=
		\begin{cases}
		\frac{w^{0}\left(\rho_{j}-\rho_{cr}^{t}\right)}{\rho_{cr}^{t}} & \text{for } \rho(x,t) \leq \rho_{cr}^{t} \\
		w^{0}\left(\frac{\rho_{j}-\rho(x,t)}{\rho(x,t)}\right) & \text{otherwise }
		\end{cases}
		\label{eqn:Mspeedrhocr}
		\end{equation}
	\end{spacing}
	
	In this article, we use $u_{f}^{t}$ in the estimation procedure and update $\rho_{cr}^{t}$ based on corresponding $u_{f}^{t}$ updates. In other words, we use Equation \ref{eqn:Mspeed} to generated the model predicted speed for $u_{f}^{t}$ ensembles. Then, when $u_{f}^{t}$ is updated, we determine $\rho_{cr}^{t}$ updates by substituting the ensemble mean (best $u_{f}^{t}$ estimate) in Equation \ref{eqn:ufrhoRelSimple}.\newline
	
	Following \cite{wang2005b, tampere2007}, we use a random walk to specify the forward model $\mathbf{F}$ as shown in Equation \ref{eqn:vRW}, where $z\sim\text{N}(\mathbf{0},\mathbf{Q_{u_{f}}})$ is Gaussian white noise. Then, we define the $M_{u_{f}}\times N$ matrix $\mathbf{A}$ as in Equation \ref{eqn:newAuf} and the matrix $\mathbf{\hat{A}}$ as in Equation \ref{eqn:newAhatuf} (i.e., $\mathbf{\hat{A}}$ uses Equation \ref{eqn:Mspeed}). Then, Equations \ref{eqn:vRW}--\ref{eqn:lastParamNew} represent the EnKF for propagating and updating free flow speed estimates when speed measurements are obtained.

	\begin{spacing}{1.25}
		\begin{equation}
		\mathbf{u}_{f, i}^{t+\Delta t|t} = \mathbf{u}_{f, i}^{t} + z
		\label{eqn:vRW}
		\end{equation}
		\begin{equation}
		\mathbf{A}=\left[\mathbf{u}_{f, 1}^{t+\Delta t|t}, \mathbf{u}_{f, 2}^{t+\Delta t|t}, ...,\mathbf{u}_{f, N}^{t+\Delta t|t}\right]
		\label{eqn:newAuf}
		\end{equation}
		\begin{equation}
		\mathbf{\hat{A}}=\left[\mathsf{M}(\mathbf{u}_{f, 1}^{t}), \mathsf{M}(\mathbf{u}_{f, 2}^{t}), ...,\mathsf{M}(\mathbf{u}_{f, N}^{t})\right]
		\label{eqn:newAhatuf}
		\end{equation}
		\begin{equation}
		\mathbf{\bar{A}}=\mathbf{A}\mathbf{1_{N}}
		\end{equation}
		\begin{equation}
		\mathbf{d_{j}} = \mathbf{d}^{v} + \epsilon_{j}^{v}  \quad \forall j \in \{1,..,N \}
		\end{equation}
		\begin{equation}
		\mathbf{D}=\left[\mathbf{d_{1}}, \mathbf{d_{2}}, ...,\mathbf{d_{N}}\right]
		\end{equation}
		\begin{equation}
		\Upsilon=\left[\epsilon_{1}^{v}, \epsilon_{2}^{v}, ...,\epsilon_{N}^{v}\right]
		\end{equation}
		\begin{equation}
		\mathbf{A^{a}}=\mathbf{A}+(\mathbf{A}-\mathbf{\bar{A}})(\mathbf{\hat{A}}-\mathbf{\hat{A}}\mathbf{1_{N}})^\top((\mathbf{\hat{A}}-\mathbf{\hat{A}}\mathbf{1_{N}})(\mathbf{\hat{A}}-\mathbf{\hat{A}}\mathbf{1_{N}})^\top + \Upsilon \Upsilon^\top   )^{-1}(\mathbf{D}-\mathbf{\hat{A}})\label{eqn:updateuf}
		\end{equation}
		\begin{equation}
		\mathbf{P}= \frac{1}{N-1} (\mathbf{A^{a}}-\mathbf{A^{a}}\mathbf{1_{N}})(\mathbf{A^{a}}-\mathbf{A^{a}}\mathbf{1_{N}})^\top
		\label{eqn:lastParamNew}
		\end{equation}
	\end{spacing}

	\subsubsection{The Dual State EnKF Algorithm}
	To summarize, we propagate separate filters for the traffic densities and incident parameters, with the filters recursively feeding best estimates into each other. Given traffic densities (best estimates) assimilated through the traffic densities EnKF, we update the parameters when speed measurements are observed. To maintain a monotonic relationship between parameters and observed speed measurements, we choose to estimate $u_{f}$ via Equations \ref{eqn:vRW}--\ref{eqn:lastParamNew}. Then, we update $\rho_{cr}$ at each incident prone location based on corresponding $u_{f}$ updates using Equation \ref{eqn:ufrhoRelSimple}. After the parameters are updated, we modify the traffic state EnKF forward model $\mathsf{CTM_{\Delta t}}$ using the new parameter values. Then, we update traffic densities through Equations \ref{eqn:firstTraffic}--\ref{eqn:lastTraffic} until speed measurements are observed again. Thus, the dual filters recursively feed best state estimates into each other. This dual EnKF estimation procedure is shown in Algorithm \ref{algo:DEnKF}.
	
	\begin{spacing}{1.25}
		\begin{algorithm}[H]
			\caption{Dual EnKF for traffic densities and free flow speed parameters at incident prone locations}
			\label{algo:DEnKF}
			\begin{algorithmic}[]
				\State \textbf{Initialization:}
				\State (1) Define $\mathsf{CTM_{\Delta t}}$ based on incident-free calibrated parameters
				\State (2) Create initial ensembles for densities across cells
				\State (3) Create initial ensembles for free flow speeds at incident prone locations
				\vspace{3mm}
				\State \textbf{Dual EnKF:} 
				\For {$time$ in $estimation$ $horizon$}
				\State (4) Propagate and update density ensembles using Equations \ref{eqn:firstTraffic}--\ref{eqn:lastTraffic}
				\If {get speed observation}
				\State (5) Propagate and update free flow speeds ensembles using Equations \ref{eqn:vRW}--\ref{eqn:lastParamNew}
				\State (6) Update critical density estimates using Equation \ref{eqn:ufrhoRelSimple} and the free flow speed best estimates
				\State (7) Adjust parameters in $\mathsf{CTM_{\Delta t}}$ based on updated free flow speed and critical density estimates
				\EndIf
				\EndFor
			\end{algorithmic}
		\end{algorithm}
	\end{spacing}

	\section{UAV Navigation for Uncertainty Minimization and Traffic State-Parameter Estimation}
	\label{sec:drone}
\begin{figure}[H]
		\begin{center}
			\includegraphics[width=0.7\textwidth]{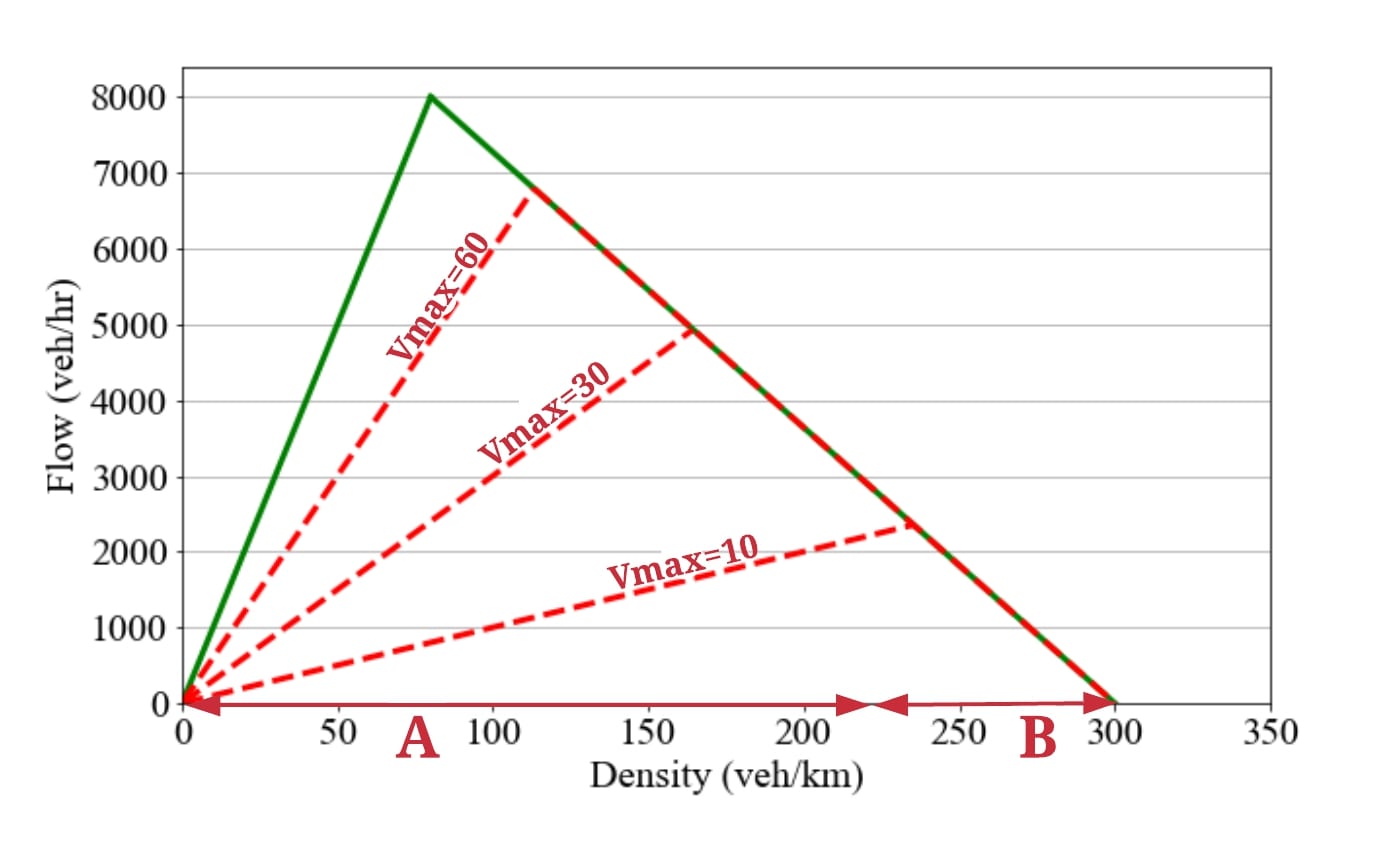}
		\end{center}
		\caption{Change in fundamental diagram with increasing incident severity using monotonic $u_{f}$, $\rho_{cr}$ updates. Dashed lines correspond to incident fundamental diagram. In region B, the fundamental diagrams can not be differentiated using speed-density observations.}
		\label{fig:FD}
	\end{figure}

	The proposed dual estimation procedure (Section \ref{sec:estim}) can efficiently estimate traffic densities and parameters in situations where the densities vary significantly during the estimation time horizon. We can also effectively infer the occurrence of incidents under uncongested conditions. In more detail, if low density measurements are accompanied with low speed measurements, then this indicates that an incident occurred. In terms of the flow-density relationship corresponding to the traffic model (illustrated in Figure \ref{fig:FD}), at low density values (region A), we can distinguish between speed observations that we expect the different fundamental diagrams to predict.\newline
	
	However, in region B of Figure \ref{fig:FD}, the fundamental diagrams coincide and can not be distinguished from speed and density observations. Specifically, given high densities and low speed measurements, we would not be able to determine whether the observations correspond to congested conditions in an incident-free fundamental diagram or if there is a reduction in physical capacity. Note that poor observability under congested conditions is not specific to the triangular fundamental diagram. For any fundamental diagram shape, if the incident fundamental diagram under congested conditions maintains a similar form to the incident-free fundamental diagram, then it will be difficult to identify the true road condition using the measured speed-density data. In terms of the parameters EnKF, poor observability impacts the error covariance matrix $\mathbf{P}$; in particular, the variance of parameter state variables that are poorly observable increases with time. \newline

	To address this estimation problem under congested conditions (region B), we propose the use of unmanned aerial vehicles to directly estimate the incident state. We consider that the UAV can collect accurate density measurements as well as $u_{f}$ observations up to observation errors. We also assume that UAV density measurements are more accurate than loop detector data. Thus, in the traffic densities EnKF (Equations \ref{eqn:firstTraffic}--\ref{eqn:lastTraffic}), the component of $\mathbf{d}^{\rho}$ at the UAV location uses UAV measurements. Similarly, the measurement error covariance matrix $\mathbf{R_{\rho}}$ reflects the UAV observation error at the UAV location.\newline
	
	As for the parameters EnKF, in addition to the parameter updates performed when speed measurements are observed (Equations \ref{eqn:vRW}--\ref{eqn:lastParamNew}), the parameters are also updated once the UAV arrives at an incident prone location through Equations \ref{eqn:vRWUAV}--\ref{eqn:lastParamNewUAV}. In Equation \ref{eqn:dUAV}, $\mathbf{d}^{u_{f}}$ is a scalar representing the UAV $u_{f}$ observation at the incident prone location, where $\epsilon^{u_{f}}\sim\text{N}(0,\sigma^{2}_{u_{f}})$ is the measurement error. In contrast to Equation \ref{eqn:updateuf}, since the state variables are directly observed, the update Equation \ref{eqn:updateEqHUAV} uses a linear observation operator $\mathbf{H}$; in this case, $\mathbf{H}$ is a $1\times M_{u_{f}}$ row vector with $1$ at the component of $\mathbf{u}_{f}$ that is observed and $0$ everywhere else.

	\begin{spacing}{1.25}
		\begin{equation}
		\mathbf{u}_{f, i}^{t+\Delta t|t} = \mathbf{u}_{f, i}^{t} + z
		\label{eqn:vRWUAV}
		\end{equation}
		\begin{equation}
		\mathbf{A}=\left[\mathbf{u}_{f, 1}^{t+\Delta t|t}, \mathbf{u}_{f, 2}^{t+\Delta t|t}, ...,\mathbf{u}_{f, N}^{t+\Delta t|t}\right]
		\end{equation}
		\begin{equation}
		\mathbf{\hat{A}}=\left[\mathsf{M}(\mathbf{u}_{f, 1}^{t}), \mathsf{M}(\mathbf{u}_{f, 2}^{t}), ...,\mathsf{M}(\mathbf{u}_{f, N}^{t})\right]
		\end{equation}
		\begin{equation}
		\mathbf{\bar{A}}=\mathbf{A}\mathbf{1_{N}}
		\end{equation}
		\begin{equation}
		\mathbf{d_{j}} = \mathbf{d}^{u_{f}} + \epsilon^{u_{f}}_{j}  \quad \forall j \in \{1,..,N \}
		\label{eqn:dUAV}
		\end{equation}
		\begin{equation}
		\mathbf{D}=\left[\mathbf{d_{1}}, \mathbf{d_{2}}, ...,\mathbf{d_{N}}\right]
		\end{equation}
		\begin{equation}
		\Upsilon=\left[\epsilon_{1}^{u_{f}}, \epsilon_{2}^{u_{f}}, ...,\epsilon_{N}^{u_{f}}\right]
		\end{equation}
		\begin{equation}
		\mathbf{A^{a}}=\mathbf{A}+(\mathbf{A}-\mathbf{\bar{A}})(\mathbf{A}-\mathbf{\bar{A}})^\top\mathbf{H}^\top(\mathbf{H}(\mathbf{A}-\mathbf{\bar{A}})(\mathbf{A}-\mathbf{\bar{A}})^\top\mathbf{H}^\top + \Upsilon \Upsilon^\top   )^{-1}(\mathbf{D}-\mathbf{H}\mathbf{A})
		\label{eqn:updateEqHUAV}
		\end{equation}
		\begin{equation}
		\mathbf{P}= \frac{1}{N-1} (\mathbf{A^{a}}-\mathbf{A^{a}}\mathbf{1_{N}})(\mathbf{A^{a}}-\mathbf{A^{a}}\mathbf{1_{N}})^\top
		\label{eqn:lastParamNewUAV}
		\end{equation}
	\end{spacing}

	\subsection{A-optimal Control Trajectory Planning Objective}
	Thus, given the additional measurements that could be collected using a UAV, we develop a path planning algorithm that navigates the UAV towards informative observations. In particular, the navigation algorithm identifies UAV paths that minimize the anticipated future variance associated with the dual EnKF traffic/parameter state estimates. To minimize the mean uncertainty in an \textit{online} setting where the state covariance matrices are continuously updated, we develop a one-step lookahead path planning algorithm. Effectively, the proposed algorithm navigates the UAV towards congested locations where it is difficult to infer the incident severity from speed-density data.  Navigating the UAV to minimize the average variance of the traffic/parameter state estimates is an instance of A-optimal control \citep{ucinski2004, atkinson2007, sim2005}. \newline
	
	Let $\psi_{p}^{t+\Delta T}$ denote the traffic or parameter state vector after $\Delta T$ \textit{future} UAV and density observations along path $p$. In addition, let  $\hat{\psi}_{p}^{t+\Delta T}$ be the corresponding vector of \textit{best} traffic or parameter state estimates (i.e., the ensemble mean). The average variance of $\psi_{p}^{t+\Delta T}$ represents the future \textit{uncertainty} in the system if path $p$ is traversed by the UAV; this average variance is denoted by $J_{p}^{t+\Delta T}$ and is defined in Equation \ref{eqn:Jp}. Observe that the uncertainty measure reduces to the trace of the future covariance matrix $\mathbf{P}^{t+\Delta T}_{p}$.

	\begin{spacing}{1.25}
		\begin{equation}
		\begin{aligned}
		J_{p}^{t+\Delta T} = & \EX[||\psi_{p}^{t+\Delta T}-\hat{\psi}_{p}^{t+\Delta T}||_{2}^{2}]= \\ & \mathsf{tr}(\EX[(\psi_{p}^{t+\Delta T}-\hat{\psi}_{p}^{t+\Delta  T})(\psi_{p}^{t+\Delta T}-\hat{\psi}_{p}^{t+\Delta T})^\top])=
		\mathsf{tr}(\mathbf{P}^{t+\Delta T}_{p})
		\end{aligned}
		\label{eqn:Jp}
		\end{equation}
	\end{spacing}

	The units chosen to represent densities/speed impact the magnitude of the resulting uncertainty measure $J_{p}^{t+\Delta T}$ \citep{sim2005}. As a result, to compare the \textit{relative} uncertainty between traffic densities and parameters, we normalize the uncertainty measure based on the scale of each state variable. Since the dual EnKF maintains separate error covariance matrices, $J_{p}^{t+\Delta T}$ can be computed separately for traffic densities and incident parameters. Then, we can define an aggregate uncertainty measure that is a weighted sum of the trace matrices, where we use the weights $\lambda\in [0,1]$ to account for the differences in scale between densities and free flow speeds. The weights can also be used to represent the importance of minimizing the uncertainty on parameters relative to traffic densities. In addition, we should further normalize the uncertainty measure based on the number of traffic state variables $K$ in the densities EnKF and the number of parameters $V$ in the free flow speeds EnKF. To be more precise, let $J_{p, \rho}^{t+\Delta T}$ denote the trace of the traffic densities covariance matrix after $\Delta T$ EnKF updates using UAV and density observations along path $p$. Furthermore, let $J_{p, u_{f}}^{t+\Delta T}$ be the corresponding measure for free flow speeds. We can specify the aggregate future uncertainty measure $J_{p}^{t+\Delta T}$ associated with path $p$ as in Equation \ref{eqn:bigJp}. Then, we aim to determine the path $p^{*}$ that minimizes this aggregate uncertainty measure among the set of $m$ candidate trajectories $\{p_{1},..,p_{m}  \}$ as shown in Equation \ref{eqn:bestp}.

	\begin{spacing}{1.25}
	\begin{equation}
	J_{p}^{t+\Delta T} = \frac{\lambda}{V} J_{p, u_{f}}^{t+\Delta T} + \frac{(1-\lambda)}{K} J_{p, \rho}^{t+\Delta T}
	\label{eqn:bigJp}
	\end{equation}
	\begin{equation}
		p^{\ast} = \mathsf{argmin}_{p_{1},..,p_{m}} J_{p}^{t+\Delta T}
	\label{eqn:bestp}
	\end{equation}
	\end{spacing}

	To compute $\hat{\psi}_{p}^{t+\Delta T}$ and $J_{p}^{t+\Delta T}$, we need to embed an EnKF that propagates ensemble members at the current time $t$ into the future $t+\Delta T$ using anticipated observations along path $p$. Therefore, the framework in Figure 1 is composed of (1) a global dual EnKF that updates state error covariance matrices at every time step using actual UAV and ground sensor measurements, and (2) multiple dual EnKFs that are initiated at every time step to propagate current ensemble members into the future based on anticipated measurements along each path.\\

	Thus, to compute $J_{p}^{t+\Delta T}$, we need to define the anticipated observations along candidate paths. We can determine the UAV location at every time step in $\Delta T$ using the UAV speed and its direction of movement along a path. We also assume that the UAV can observe a specified length of the road underneath its location. Then, for every future time step in $\Delta T$, we assume that the density observations will be equal to the mean of density ensembles propagated by the forward model. In other words, the cell transmission model $\mathsf{CTM_{\Delta t}}$ is used to propagate the density ensembles at time $t$ up to the desired time step, and the mean of the propagated ensembles is considered to be the future density measurements. Similarly, for the free flow speed observations, we propagate the current ensembles into the future using the random walk forward model and consider the ensemble mean to be the anticipated observations.\\
	
	In summary, using the anticipated observations along each path, we use the embedded EnKFs to generate the future covariance matrix $\mathbf{P}^{t+\Delta T}_{p}$, we compute the uncertainty measure $J_{p}^{t+\Delta T}$, and we determine the uncertainty minimizing UAV path $p^{\ast}$.\newline

	\subsection{Online Path Planning}
	Once the UAV moves along the variance minimizing path $p^{*}$ as determined by the trajectory planning objective in Equations \ref{eqn:bigJp} and \ref{eqn:bestp}, it feeds accurate density measurements and direct $u_{f}$ observations to the global dual state EnKF. Then, the global dual state EnKF updates the traffic/parameter state estimates, covariance matrices, and $\mathsf{CTM_{\Delta t}}$ parameters based on observations from all available sensors (loop detectors, probe vehicles, and UAV measurements). Since the network information is continuously updated, the resulting anticipated future uncertainty measure $J_{p}^{t+\Delta T}$ dynamically changes. Thus, to calculate $J_{p}^{t+\Delta T}$, the covariance matrix after $\Delta T$ time steps should be obtained by propagating the \textit{current} ensembles into the future using the embedded EnKFs. In other words, $J_{p}^{t+\Delta T}$ must be re-calculated at every update step using the updated covariance matrices, UAV position, and traffic model.\newline
	
	To control the UAV in this online setting, we use a one step lookahead path planning policy \citep{bertsekas2017}. After each update stage, the proposed policy navigates the UAV in the direction of the path that minimizes the uncertainty measure. Since the number of candidate paths grows significantly with the size of the network and the possible UAV movements, we restrict our analysis to a subset of the possible UAV paths. In particular, for path enumeration, we assume that the UAV does not hover or backtrack. The path planning and estimation framework is described in Algorithm \ref{algo:estimtion-planning}.

	\begin{spacing}{1.25}
		\begin{algorithm}[H]
			\caption{Dual EnKF for traffic densities and free flow speed parameters at incident prone locations}
			\label{algo:estimtion-planning}
			\begin{algorithmic}[]
				\State \textbf{Initialization:}
				\State (1) Define $\mathsf{CTM_{\Delta t}}$ based on incident-free calibrated parameters
				\State (2) Create initial ensembles for densities across cells
				\State (3) Create initial ensembles for free flow speeds at incident prone locations
				\State (4) Set initial UAV location
				\vspace{3mm}
				\State \textbf{Dual EnKF \& UAV navigation:} 
				\For {$time$ in $estimation$ $horizon$}
				\State (5) Propagate and update density ensembles using Equations \ref{eqn:firstTraffic}--\ref{eqn:lastTraffic} (use UAV density measurements \State at UAV location)
				\If {get speed observation}
				\State (6) Propagate and update free flow speeds ensembles using Equations \ref{eqn:vRW}--\ref{eqn:lastParamNew} 
				\State (7) Update critical density estimates using Equation \ref{eqn:ufrhoRelSimple} and the free flow speed best estimates
				\State (8) Adjust parameters in $\mathsf{CTM_{\Delta t}}$ based on updated free flow speed and critical density estimates
				\EndIf
				\If {UAV at incident prone location}
				\State (9) Propagate and update free flow speeds ensembles using Equations \ref{eqn:vRWUAV}--\ref{eqn:lastParamNewUAV}
				\State (10) Update critical density estimates using Equation \ref{eqn:ufrhoRelSimple} and the free flow speed best estimates
				\State (11) Adjust parameters in $\mathsf{CTM_{\Delta t}}$ based on updated free flow speed and critical density estimates
				\EndIf
				\State (12) Generate possible UAV paths from current location
				\State (13) For $\Delta T$ future time steps, determine UAV location at every time instant along each path
				\State (14) Generate anticipated UAV observations along candidate paths 
				\State (15) Use the embedded dual state EnKFs for $\Delta T$ steps into the future to determine $J_{p}^{t+\Delta T}$ along each \State candidate path
				\State (16) Move the UAV in the direction of $p^{*}$ (the path that minimizes  $J_{p}^{t+\Delta T}$) 
				\EndFor
			\end{algorithmic}
		\end{algorithm}
	\end{spacing}

	\section{Results}
	\label{sec:results}
\begin{figure}
		\begin{center}
			\includegraphics[width=0.8\textwidth]{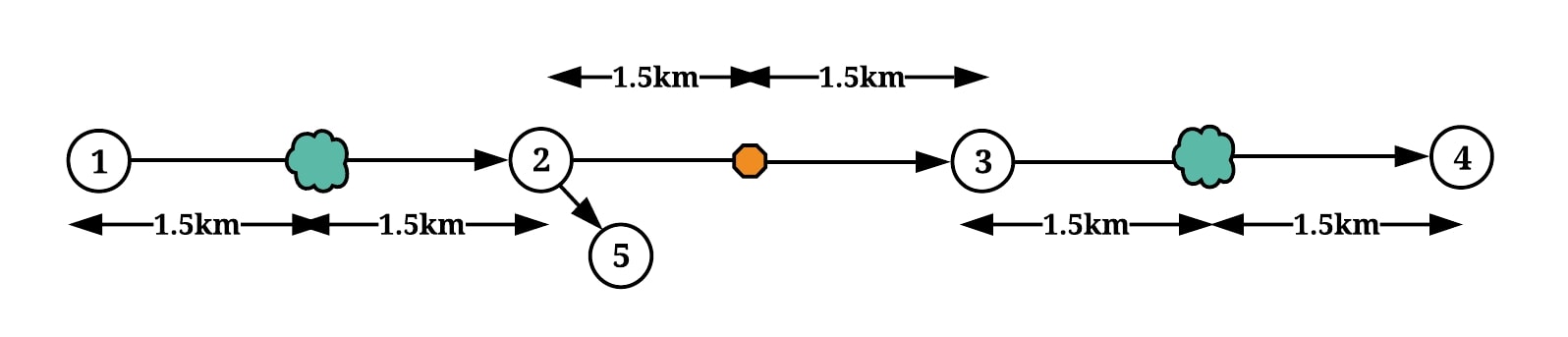}
		\end{center}
		\caption{VISSIM network.}
		\label{fig:net}
	\end{figure}

	To demonstrate the benefit of targeted observation in improving incident detection and traffic state estimation, we implement the proposed algorithms on a freeway with an off-ramp modeled in VISSIM. For comparison, we also implement the California algorithm (a data-driven approach developed by \cite{payne1976} for detecting incidents). The freeway length, UAV starting position, and incident prone locations are shown in Figure \ref{fig:net}. In this figure, clouds indicate incident locations and the middle circle indicates the UAV starting position.  \newline

	Field data collected by \cite{pan2013} and \cite{quiroga2004} suggests that the maximum speed under incident conditions is in the range of $15$--$30$ km/hr. Thus, to model non-recurrent congestion at incident prone locations in VISSIM, we specify a reduced speed zone where the maximum speed is set at $20$ km/hr. To simulate congested incident conditions upstream (region B in Figure \ref{fig:FD}) and uncongested incident conditions downstream (region A in Figure \ref{fig:FD}), we consider that at node 2 half of the inflow demand continues on to node 4 while the other half takes the off-ramp. \newline
	
	For the traffic densities EnKF (Equations \ref{eqn:firstTraffic}--\ref{eqn:lastTraffic}), we assume that $\mathbf{Q_{\rho}}$ is a diagonal matrix such that the diagonal entries are all $(5 \text{ veh/km})^{2}$. We also assume that $\mathbf{R_{\rho}}$ is a diagonal matrix with elements $(10 \text{ veh/km})^{2}$. However, when a traffic density state variable is observed by a UAV, the corresponding component in $\mathbf{R_{\rho}}$ is $(2 \text{ veh/km})^{2}$. For the incident parameters EnKF, we consider that $\mathbf{Q_{u_{f}}}$ is a diagonal matrix with entries $(5 \text{ km/hr})^{2}$. When speed measurements are observed (Equations \ref{eqn:vRW}--\ref{eqn:lastParamNew}), the measurement error covariance matrix $\mathbf{R_{v}}$ is a diagonal matrix with entries $(5 \text{ km/hr})^{2}$. When a UAV directly observes incident conditions (Equations \ref{eqn:vRWUAV}--\ref{eqn:lastParamNewUAV}), the measurement error $\sigma^{2}_{u_{f}}$ is $(10 \text{ km/hr})^{2}$. The number of ensembles $N$ in each EnKF is $100$. We consider that loop detectors feed density measurements at every time step ($10$ seconds), and that speed measurements are collected from GPS equipped probe vehicles every $30$ time steps ($5$ minutes). The initial calibrated model parameters in the estimation procedure are as follows: $\rho_{cr}^{0}=80\text{ veh/km}$, $u_{f}^{0}=100\text{ km/hr}$, $\rho_{j}=300\text{ veh/km}$. \newline

	For UAV path planning (Algorithm \ref{algo:estimtion-planning}), we set $\lambda = 0.5$ to represent equal weights for traffic densities and parameter state uncertainty measures. In Section \ref{sec:results:drone}, we further study the impact of the weight $\lambda$ on the UAV trajectory. We determine $\Delta T$ dynamically as the number of time steps until the UAV reaches node 1 if it is traveling upstream, or the number of time steps until it reaches node 4 if it is traveling downstream. In addition, we assume that the UAV can observe $250$ meters at every time step. \newline

	\subsection{Traffic State \& Incident Detection Estimation Results}
	\label{sec:results:estim}
\begin{figure}[H]
		\begin{center}
			\includegraphics[width=0.9\textwidth]{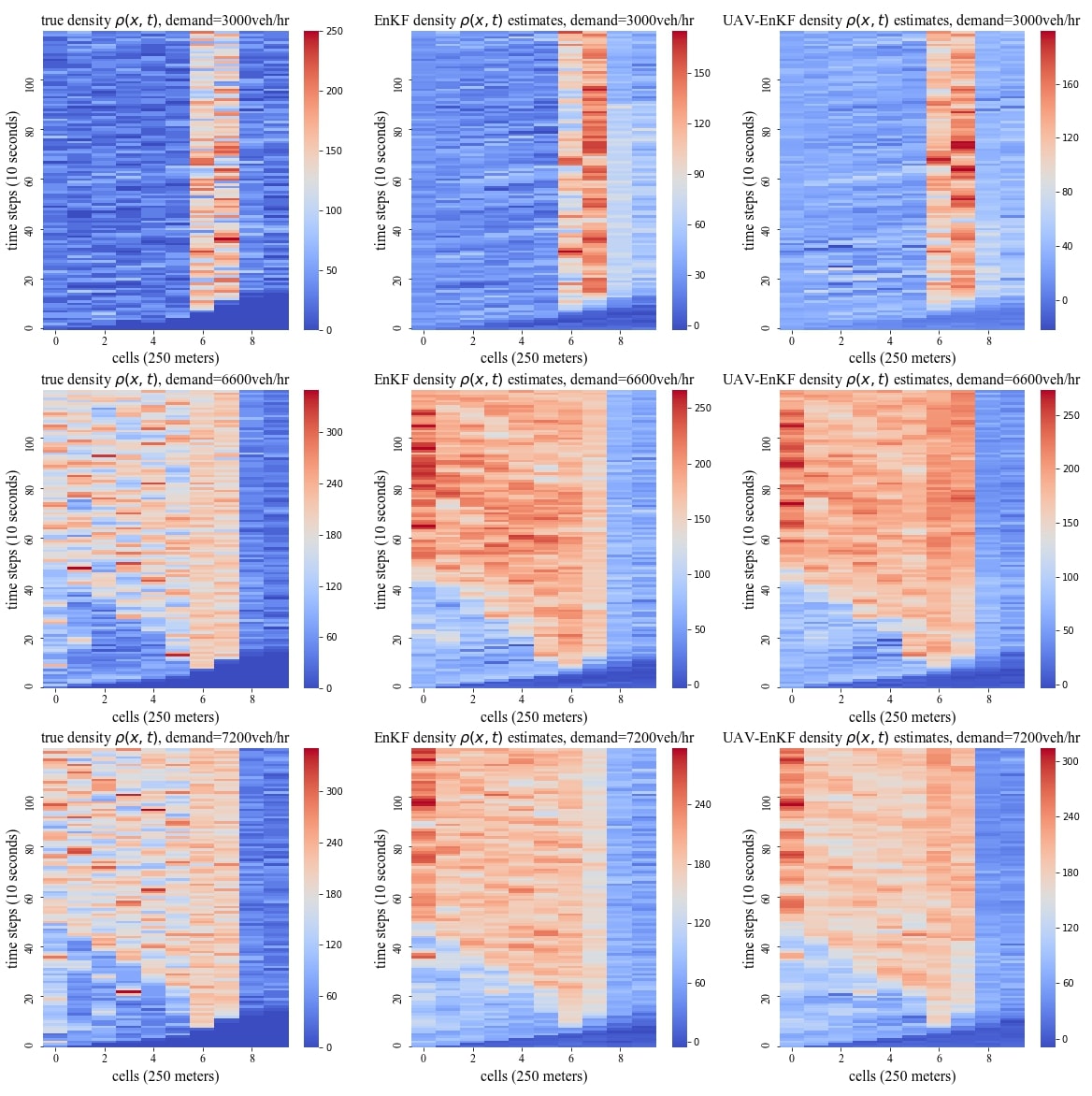}
		\end{center}
		\caption{True density values, EnKF density estimates (Algorithm \ref{algo:DEnKF}), and UAV-EnKF density estimates (Algorithm \ref{algo:estimtion-planning}) at the upstream incident location (cells 6 and 7).}
		\label{fig:den}
	\end{figure}

	We evaluate the performance of the proposed estimation (EnKF) and estimation-planning (UAV-EnKF) algorithms under three different levels of inflow (at node 1) demand: 3000veh/hr, 6600veh/hr, and 7200veh/hr. Figure \ref{fig:den} illustrates the density estimates relative to the true simulation data at the upstream incident location. In this figure, the links are discretized into cells that are used by the cell transmission model, and the upstream incident occurs at cells 6 and 7. As shown in Figure \ref{fig:den}, since densities are directly observed by ground sensors, they are accurately estimated by both EnKF and UAV-EnKF algorithm.
	
\begin{figure}[H]
		\begin{center}
			\includegraphics[width=0.9\textwidth]{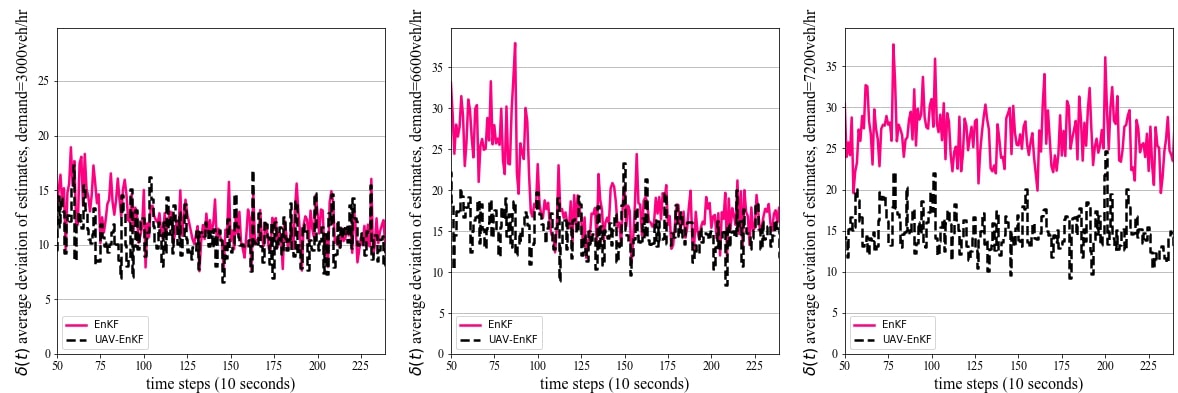}
		\end{center}
		\caption{Average of the absolute differences between the estimates and true density values.}
		\label{fig:dev}
	\end{figure}

	To further analyze the density estimates, we define $\delta(t)$ as the average deviation of estimates from the true density values. Precisely, $\delta(t)$ is defined as the average (across cells $\{0,..,n\}$ at time $t$) of the absolute differences between the estimates and true densities. In Equation \ref{eqn:deviation}, $\hat{\rho}(x,t)$ is the density estimate at cell $x$ and time $t$ and $\bar{\rho}(x,t)$ is the true density value at cell $x$ and time $t$. In Figure \ref{fig:dev}, we illustrate the variation in $\delta(t)$ across time for the EnKF and UAV-EnKF algorithms. We observe that the UAV-EnKF estimates are better than the corresponding EnKF estimates as indicated by a lower average absolute deviation $\delta(t)$.
	
	\begin{spacing}{1.25}
		\begin{equation}
		\delta(t) = \frac{1}{n}\sum_{x=0}^{n}|\hat{\rho}(x,t) - \bar{\rho}(x,t)|
		\label{eqn:deviation}
		\end{equation}
	\end{spacing}
	
\begin{figure}[H]
		\begin{center}
			\includegraphics[width=0.9\textwidth]{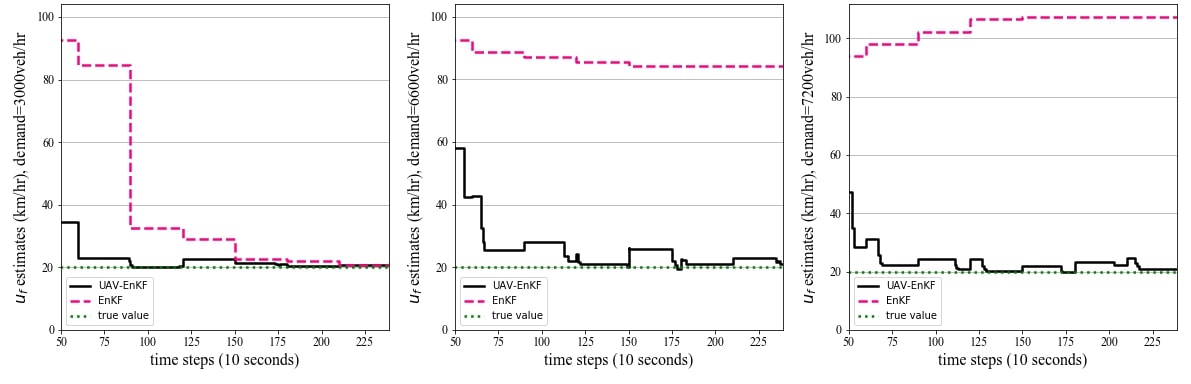}
		\end{center}
		\caption{Free flow speed $u_{f}$ estimates generated by the EnKF and UAV-EnKF algorithms at the upstream incident location.}
		\label{fig:vmaxUP}
	\end{figure}

\begin{figure}[H]
		\begin{center}
			\includegraphics[width=0.9\textwidth]{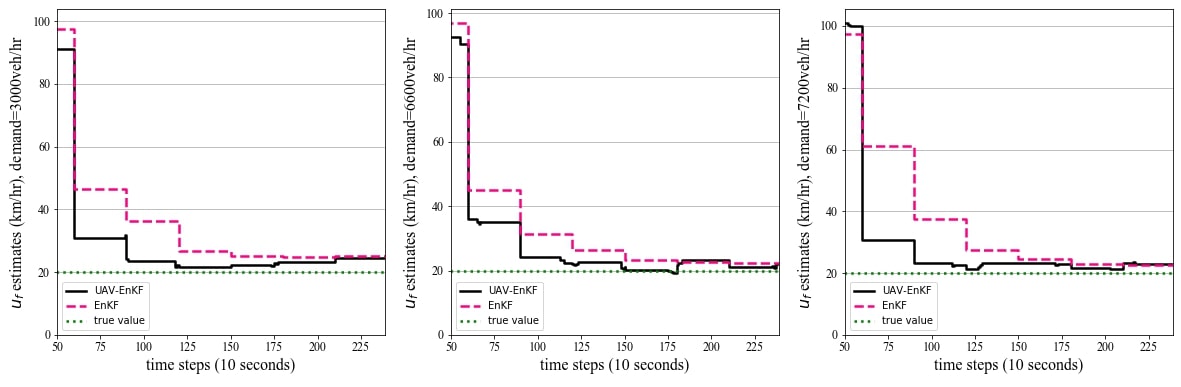}
		\end{center}
		\caption{Free flow speed $u_{f}$ estimates generated by the EnKF and UAV-EnKF algorithms at the downstream incident location.}
		\label{fig:vmaxDW}
	\end{figure}

\begin{figure}[H]
		\begin{center}
			\includegraphics[width=0.9\textwidth]{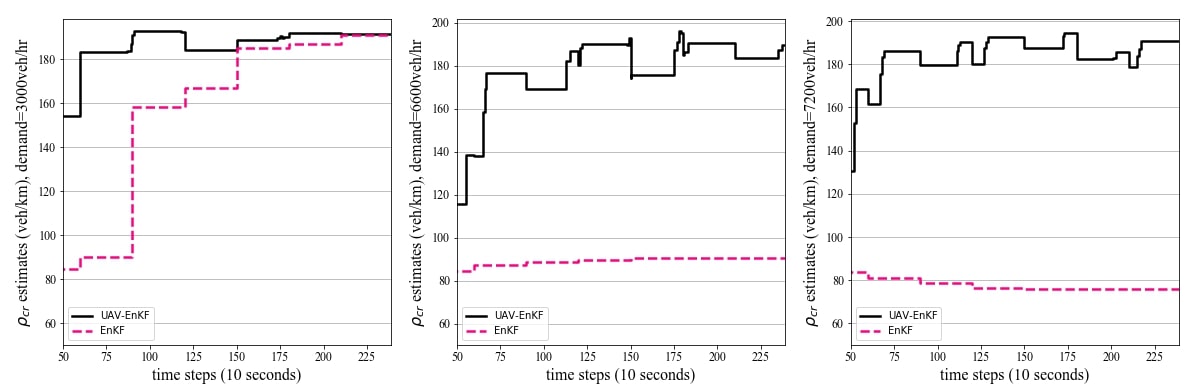}
		\end{center}
		\caption{Critical density $\rho_{cr}$ estimates generated by the EnKF and UAV-EnKF algorithms at the upstream incident location.}
		\label{fig:rhoC}
	\end{figure}

	In terms of model parameters representing incidents ($u_{f}$ and $\rho_{cr}$), we illustrate in Figure \ref{fig:vmaxUP} (upstream incident location) and Figure \ref{fig:vmaxDW} (downstream incident location) the free flow speed $u_{f}$ estimates. We also illustrate in Figure \ref{fig:rhoC} the corresponding critical density updates at the upstream incident location, where the critical density at any stage is updated according to Equation \ref{eqn:ufrhoRelSimple}.


	As shown in Figures \ref{fig:vmaxUP} and \ref{fig:rhoC}, under congested conditions (demand=$6600$veh/hr and $7200$veh/hr) at the upstream incident location, the EnKF estimates do not represent the true incident conditions. The poor $u_{f}$ and $\rho_{cr}$ estimates are caused by the similarity between incident and congested incident-free measurements (high density and low speed measurements); in particular, this similarity leads to poor observability as previously discussed in Section \ref{sec:drone}. On the other hand, when the inflow demand is low, the EnKF accurately determines the true $u_{f}$ and $\rho_{cr}$ parameters.\newline

	To analyze the incident detection performance further, we implement the California data-driven incident detection algorithm. The California algorithm was proposed by \cite{payne1976} and further analyzed by \cite{wang2016, stephanedes1993, parkany2005}. The algorithm applies three different tests that compare occupancy data upstream and downstream of the incident prone location to predefined thresholds. In our implementation, we use the same thresholds that \cite{wang2016} used (T1=0.27, T2=0.55, and T3=0.0003). \cite{payne1976, parkany2005} provide additional details on algorithm implementation. As shown in Table \ref{tab:matable}, the California algorithm was not able to detect the incidents under low/intermediate flow conditions ($\leq 3600$veh/hr). This failure in detection is caused by the lack of propagation of incident information. In other words, since the traffic flow is relatively low, queues do not build up towards the sensor upstream of the incident; thus, the algorithm is not capable of detecting a significant difference in occupancy between sensor measurements upstream and downstream of the incident.

	\begin{table}[H]
		\fontsize{10pt}{12pt}\selectfont
		\centering
		\caption{Incident detection performance for California, EnKF, and UAV-EnKF algorithms.}
		\begin{tabular}{|
				>{\columncolor[HTML]{EFEFEF}}l |l|l|l|l|l|l|}
			\hline
			{\color[HTML]{000000} } & \multicolumn{2}{|l|}{\cellcolor[HTML]{EFEFEF}{\color[HTML]{000000} 3000 veh/hr}} & \multicolumn{2}{|l|}{\cellcolor[HTML]{EFEFEF}{\color[HTML]{000000} 6600 veh/hr}} & \multicolumn{2}{|l|}{\cellcolor[HTML]{EFEFEF}{\color[HTML]{000000} 7200 veh/hr}} \\ \cline{2-7}
			& upstream                              & downstream                              & upstream                              & downstream                              & upstream                              & downstream                              \\ \cline{2-7}
			California              &        not detected                                &                                 not detected       &         detected                              &        not detected                                 &            detected                           &                     not detected                    \\
			EnKF                    &           detected                            &    detected                                     &       not detected                                &           detected                              &         not detected                              &        detected                                 \\
			UAV-EnKF                &      detected                                 &                   detected                      &         detected                              &                        detected                 &        detected                               &              detected  \\ \hline                        
		\end{tabular}
	\label{tab:matable}
	\end{table}

	The inability of the California algorithm to detect incidents when the demand is low and the inability of the EnKF algorithm to detect incidents when the demand is high can be considered as Type II errors. In this case, Type II errors represent rejecting the hypothesis that there are no incidents even though an incident exists. However, comparative data-driven methods such as the California algorithm may also results in Type I errors (reporting an incident under incident-free conditions) due to random fluctuations in the data \citep{stephanedes1993, parkany2005}. In contrast, model based filtering algorithms such as the EnKF incorporate sensor noise in the estimation procedure, where the extent of correction to match the measurements and the associated variance of the estimates depends on the magnitude of the sensor noise relative to the model noise. As for the UAV-EnKF approach, we assume that the incident condition can be accurately and instantaneously detected by UAV images once the UAV arrives at the incident location. Similar to our work, \cite{wang2016} studies the incident detection performance in simulation across varying congestion levels, and they also report that the California algorithm fails to detect incidents when the demand is low ($\leq4000$ veh/hr).

	\subsection{UAV Path Planning}
	\label{sec:results:drone}
\begin{figure}[H]
		\begin{center}
			\includegraphics[width=0.9\textwidth]{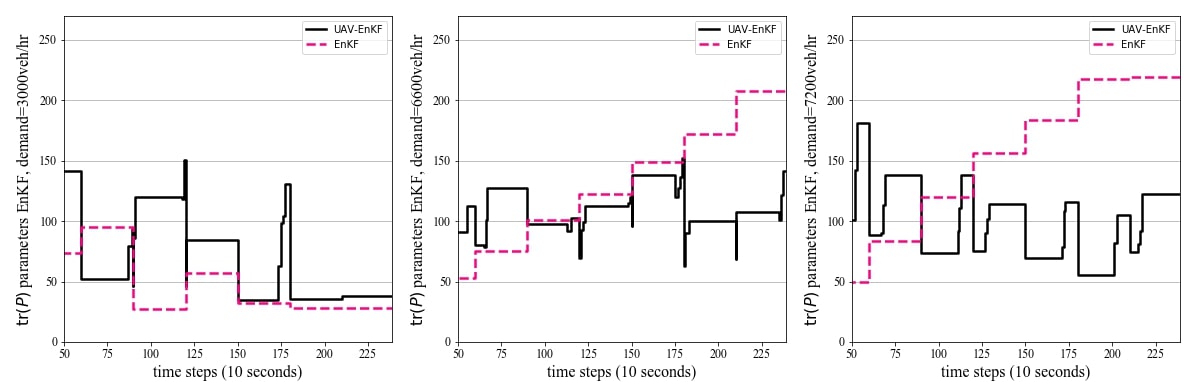}
		\end{center}
		\caption{Trace of the $u_{f}$ covariance matrix associated with the EnKF estimates at the upstream incident location. $\lambda=0.5$.}
		\label{fig:trP}
	\end{figure}

	The UAV trajectory is determined by the average variance on density and parameter estimates as defined by Equations \ref{eqn:Jp} and \ref{eqn:bigJp}. Recall that we expect the variance on the EnKF parameter estimates to have an increasing trend under congested conditions (region B of Figure \ref{fig:FD}), where this increasing trend results from poor observability. To illustrate, we show in Figure \ref{fig:trP} the trace of the $u_{f}$ covariance matrix at the upstream incident location. We observe that under congested conditions, the average variance $\mathsf{tr}(\mathbf{P})$ associated with the EnKF algorithm is increasing with time. Another indication of the increase in average variance under congested conditions is the smoothness of the parameter estimates. As shown in Figure \ref{fig:vmaxUP}, when the inflow demand is $3000$veh/hr, the parameter estimates are smoother than the corresponding estimates under congested conditions (demand=$6600$ or $7200$veh/hr).
	
\begin{figure}[H]
		\begin{center}
			\includegraphics[width=0.9\textwidth]{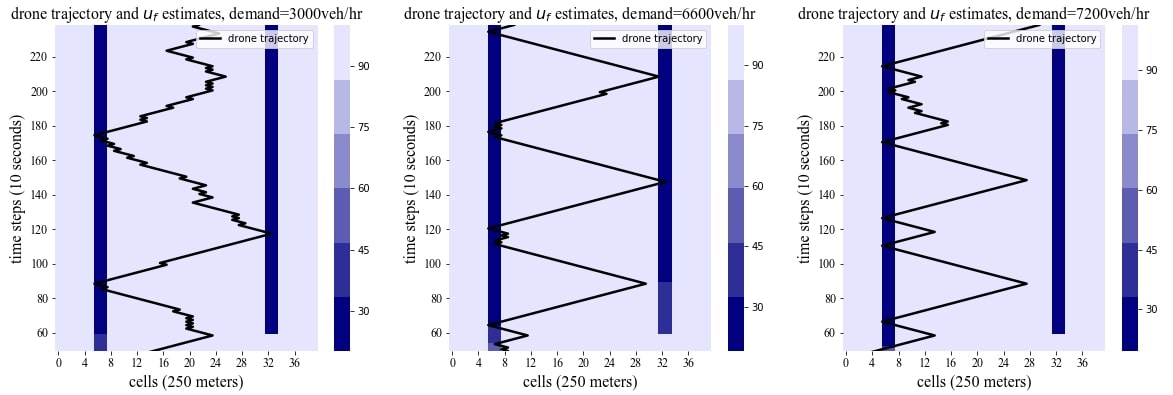}
		\end{center}
		\caption{UAV trajectory for different levels of inflow demand. The heatmap corresponds to $u_{f}$ estimates. $\lambda=0.5$.}
		\label{fig:drtrajC}
	\end{figure}

	To further illustrate the impact of uncertainty on the UAV trajectory, we show in Figure \ref{fig:drtrajC} the UAV path for varying inflow demand. Since the traffic density is higher at the upstream incident relative to the downstream incident, we observe that the UAV spends more time at the upstream incident location (the average variance on parameter estimates would be higher upstream). In addition, as the inflow demand increases, we observe that the UAV will spend relatively more time at the increasingly congested upstream incident location.
	
\begin{figure}[H]
		\begin{center}
			\includegraphics[width=0.77\textwidth]{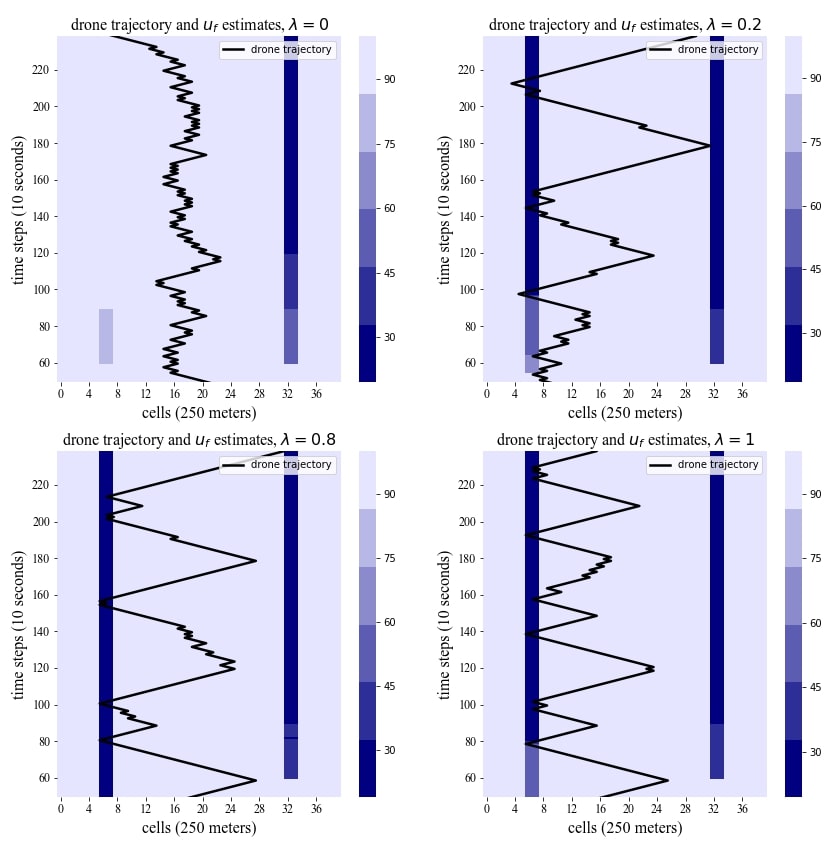}
		\end{center}
		\caption{UAV trajectory for different values of the mean uncertainty weighting factor $\lambda$. The heatmap corresponds to $u_{f}$ estimates. Inflow demand=$6600$veh/hr.}
		\label{fig:drtrajWeights}
	\end{figure}

	We also study the impact of the weighting factor $\lambda$ in Equation \ref{eqn:bigJp} on the UAV trajectory, where $\lambda$ weights the average variance on the parameter estimates relative to the average variance on the traffic density estimates. Figure \ref{fig:drtrajWeights} illustrates the UAV trajectory and parameter estimates for different values of $\lambda$ when the demand is $6600$veh/hr ($\lambda=0.5$ is shown in Figure \ref{fig:drtrajC}). As observed, when $\lambda=0$, indicating that the path planning objective is dominated by the uncertainty on the density estimates, the UAV does not navigate towards the incident locations; in turn, this results in poor parameter estimates at the upstream incident location. When $\lambda>0$, we observe that the UAV navigates towards the incident locations and that the corresponding parameter estimates reflect the true incident severity. As $\lambda$ increases (i.e., the weight on the variance of parameter estimates increases), we observe that the UAV spends more time closer to the congested upstream incident location.

	\section{Conclusion}
	\label{sec:conc}
	Non-recurrent congestion is a primary source of travel time variability and congestion delays. Traditional data-driven methods for detecting incidents are susceptible to false alarms. In addition, data-driven methods lack a traffic model for predicting the congestion state beyond the incident location. On the other hand, methods that use a macroscopic traffic model to simultaneously estimate traffic conditions and incident severity suffer from poor observability in congested conditions; in particular, it is difficult to distinguish speed/density measurements under incident-free congested conditions from similar observations due to incidents.\newline
	
	We propose a planning-estimation framework that relies on unmanned aerial vehicles (UAVs) to generate targeted observations. Specifically, we develop an online one-step lookahead algorithm that uses a dual ensemble Kalman filter (EnKF) to determine the uncertainty minimizing UAV path. In the dual EnKF estimation procedure, we implement a parameter update technique that maintains a monotonic relationship between observed measurements and parameters. We test the UAV planning-estimation framework on a freeway segment and compare its performance against methods that do not use targeted incident observations. We observe that the UAV observations aid in detecting the road condition when it is otherwise difficult to infer the road state from the speed-density data. 
	
	\section*{Supplementary Material}
	Data and code are available on \url{https://github.com/spartalab/uav-path-plan}.
	
	\section*{Acknowledgments}
	The authors gratefully acknowledge the support of the Center for Advanced Multimodal Mobility Solutions and Education (CAMMSE) and the National Science Foundation under Grant No. 1636154.


{ \footnotesize
	\begin{thebibliography}{36}
\expandafter\ifx\csname natexlab\endcsname\relax\def\natexlab#1{#1}\fi
\providecommand{\url}[1]{\texttt{#1}}
\providecommand{\href}[2]{#2}
\providecommand{\path}[1]{#1}
\providecommand{\DOIprefix}{doi:}
\providecommand{\ArXivprefix}{arXiv:}
\providecommand{\URLprefix}{URL: }
\providecommand{\Pubmedprefix}{pmid:}
\providecommand{\doi}[1]{\href{http://dx.doi.org/#1}{\path{#1}}}
\providecommand{\Pubmed}[1]{\href{pmid:#1}{\path{#1}}}
\providecommand{\bibinfo}[2]{#2}
\ifx\xfnm\relax \def\xfnm[#1]{\unskip,\space#1}\fi
\bibitem[{{Abu-Jbara} et~al.(2015){Abu-Jbara}, Alheadary, Sundaramorthi and
	Claudel}]{abu-jbara2015}
\bibinfo{author}{{Abu-Jbara}, K.}, \bibinfo{author}{Alheadary, W.},
\bibinfo{author}{Sundaramorthi, G.}, \bibinfo{author}{Claudel, C.},
\bibinfo{year}{2015}.
\newblock \bibinfo{title}{A robust vision-based runway detection and tracking
	algorithm for automatic {{UAV}} landing}, in: \bibinfo{booktitle}{2015
	{{International Conference}} on {{Unmanned Aircraft Systems}} ({{ICUAS}})},
\bibinfo{publisher}{{IEEE}}. pp. \bibinfo{pages}{1148--1157}.
\bibitem[{Anbaroglu et~al.(2014)Anbaroglu, Heydecker and Cheng}]{anbaroglu2014}
\bibinfo{author}{Anbaroglu, B.}, \bibinfo{author}{Heydecker, B.},
\bibinfo{author}{Cheng, T.}, \bibinfo{year}{2014}.
\newblock \bibinfo{title}{Spatio-temporal clustering for non-recurrent traffic
	congestion detection on urban road networks}.
\newblock \bibinfo{journal}{Transportation Research Part C: Emerging
	Technologies} \bibinfo{volume}{48}, \bibinfo{pages}{47--65}.
\bibitem[{Atkinson et~al.(2007)Atkinson, Donev and Tobias}]{atkinson2007}
\bibinfo{author}{Atkinson, A.}, \bibinfo{author}{Donev, A.},
\bibinfo{author}{Tobias, R.}, \bibinfo{year}{2007}.
\newblock \bibinfo{title}{Optimum Experimental Designs, with {{SAS}}}.
volume~\bibinfo{volume}{34}.
\newblock \bibinfo{publisher}{{Oxford University Press}}.
\bibitem[{Barmpounakis et~al.(2019)Barmpounakis, Vlahogianni, Golias and
	Babinec}]{barmpounakis2019}
\bibinfo{author}{Barmpounakis, E.}, \bibinfo{author}{Vlahogianni, E.},
\bibinfo{author}{Golias, J.}, \bibinfo{author}{Babinec, A.},
\bibinfo{year}{2019}.
\newblock \bibinfo{title}{How accurate are small drones for measuring
	microscopic traffic parameters?}
\newblock \bibinfo{journal}{Transportation Letters} \bibinfo{volume}{11},
\bibinfo{pages}{332--340}.
\bibitem[{Bazi and Melgani(2018)}]{bazi2018}
\bibinfo{author}{Bazi, Y.}, \bibinfo{author}{Melgani, F.},
\bibinfo{year}{2018}.
\newblock \bibinfo{title}{Convolutional {{SVM}} networks for object detection
	in {{UAV}} imagery}.
\newblock \bibinfo{journal}{IEEE Transactions on Geoscience and Remote Sensing}
\bibinfo{volume}{56}, \bibinfo{pages}{3107--3118}.
\bibitem[{Bertsekas(2017)}]{bertsekas2017}
\bibinfo{author}{Bertsekas, D.P.}, \bibinfo{year}{2017}.
\newblock \bibinfo{title}{Dynamic Programming and Optimal Control}.
volume~\bibinfo{volume}{1}.
\newblock \bibinfo{edition}{Fourth} ed., \bibinfo{publisher}{{Athena
		Scientific}}.
\bibitem[{Blandin et~al.(2012)Blandin, Couque, Bayen and Work}]{blandin2012}
\bibinfo{author}{Blandin, S.}, \bibinfo{author}{Couque, A.},
\bibinfo{author}{Bayen, A.}, \bibinfo{author}{Work, D.},
\bibinfo{year}{2012}.
\newblock \bibinfo{title}{On sequential data assimilation for scalar
	macroscopic traffic flow models}.
\newblock \bibinfo{journal}{Physica D: Nonlinear Phenomena}
\bibinfo{volume}{241}, \bibinfo{pages}{1421--1440}.
\bibitem[{Canepa and Claudel(2017)}]{canepa2017}
\bibinfo{author}{Canepa, E.S.}, \bibinfo{author}{Claudel, C.G.},
\bibinfo{year}{2017}.
\newblock \bibinfo{title}{Networked traffic state estimation involving mixed
	fixed-mobile sensor data using {{Hamilton}}-{{Jacobi}} equations}.
\newblock \bibinfo{journal}{Transportation Research Part B: Methodological}
\bibinfo{volume}{104}, \bibinfo{pages}{686--709}.
\bibitem[{Cheu and Ritchie(1995)}]{cheu1995}
\bibinfo{author}{Cheu, R.L.}, \bibinfo{author}{Ritchie, S.G.},
\bibinfo{year}{1995}.
\newblock \bibinfo{title}{Automated detection of lane-blocking freeway
	incidents using artificial neural networks}.
\newblock \bibinfo{journal}{Transportation Research Part C: Emerging
	Technologies} \bibinfo{volume}{3}, \bibinfo{pages}{371--388}.
\bibitem[{Dabiri and Kulcs{\'a}r(2015)}]{dabiri2015}
\bibinfo{author}{Dabiri, A.}, \bibinfo{author}{Kulcs{\'a}r, B.},
\bibinfo{year}{2015}.
\newblock \bibinfo{title}{Freeway traffic incident reconstruction \textendash{}
	a bi-parameter approach}.
\newblock \bibinfo{journal}{Transportation Research Part C: Emerging
	Technologies} \bibinfo{volume}{58}, \bibinfo{pages}{585--597}.
\bibitem[{Daganzo(1994)}]{daganzo1994a}
\bibinfo{author}{Daganzo, C.F.}, \bibinfo{year}{1994}.
\newblock \bibinfo{title}{The cell transmission model: {{A}} dynamic
	representation of highway traffic consistent with the hydrodynamic theory}.
\newblock \bibinfo{journal}{Transportation Research Part B: Methodological}
\bibinfo{volume}{28}, \bibinfo{pages}{269--287}.
\bibitem[{Daganzo(1995)}]{daganzo1995a}
\bibinfo{author}{Daganzo, C.F.}, \bibinfo{year}{1995}.
\newblock \bibinfo{title}{The cell transmission model, part {{II}}: Network
	traffic}.
\newblock \bibinfo{journal}{Transportation Research Part B: Methodological}
\bibinfo{volume}{29}, \bibinfo{pages}{79--93}.
\bibitem[{Dia and Rose(1997)}]{dia1997}
\bibinfo{author}{Dia, H.}, \bibinfo{author}{Rose, G.}, \bibinfo{year}{1997}.
\newblock \bibinfo{title}{Development and evaluation of neural network freeway
	incident detection models using field data}.
\newblock \bibinfo{journal}{Transportation Research Part C: Emerging
	Technologies} \bibinfo{volume}{5}, \bibinfo{pages}{313--331}.
\bibitem[{Evensen(2003)}]{evensen2003}
\bibinfo{author}{Evensen, G.}, \bibinfo{year}{2003}.
\newblock \bibinfo{title}{The ensemble {{Kalman}} filter: {{Theoretical}}
	formulation and practical implementation}.
\newblock \bibinfo{journal}{Ocean dynamics} \bibinfo{volume}{53},
\bibinfo{pages}{343--367}.
\bibitem[{Evensen(2009)}]{evensen2009}
\bibinfo{author}{Evensen, G.}, \bibinfo{year}{2009}.
\newblock \bibinfo{title}{Data Assimilation: {{The}} Ensemble {{Kalman}}
	Filter}.
\newblock \bibinfo{publisher}{{Springer Science \& Business Media}}.
\bibitem[{Godunov(1959)}]{godunov1959}
\bibinfo{author}{Godunov, S.K.}, \bibinfo{year}{1959}.
\newblock \bibinfo{title}{A difference method for numerical calculation of
	discontinuous solutions of the equations of hydrodynamics}.
\newblock \bibinfo{journal}{Matematicheskii Sbornik} \bibinfo{volume}{47},
\bibinfo{pages}{271--306}.
\bibitem[{Hawas and Ahmed(2017)}]{hawas2017}
\bibinfo{author}{Hawas, Y.}, \bibinfo{author}{Ahmed, F.}, \bibinfo{year}{2017}.
\newblock \bibinfo{title}{A binary logit-based incident detection model for
	urban traffic networks}.
\newblock \bibinfo{journal}{Transportation Letters} \bibinfo{volume}{9},
\bibinfo{pages}{49--62}.
\bibitem[{Jin et~al.(2016)Jin, Ardestani, Wang and Hu}]{jin2016}
\bibinfo{author}{Jin, P.}, \bibinfo{author}{Ardestani, S.},
\bibinfo{author}{Wang, Y.}, \bibinfo{author}{Hu, W.}, \bibinfo{year}{2016}.
\newblock \bibinfo{title}{Unmanned Aerial Vehicle ({{UAV}}) Based Traffic
	Monitoring and Management}.
\newblock \bibinfo{type}{Technical Report}. {Center for Advanced Infrastructure
	and Transportation (CAIT)}.
\bibitem[{Krajewski et~al.(2018)Krajewski, Bock, Kloeker and
	Eckstein}]{krajewski2018}
\bibinfo{author}{Krajewski, R.}, \bibinfo{author}{Bock, J.},
\bibinfo{author}{Kloeker, L.}, \bibinfo{author}{Eckstein, L.},
\bibinfo{year}{2018}.
\newblock \bibinfo{title}{The {{highD}} dataset: {{A}} drone dataset of
	naturalistic vehicle trajectories on {{German}} highways for validation of
	highly automated driving systems}, in: \bibinfo{booktitle}{2018 {{IEEE}} 21st
	{{International Conference}} on {{Intelligent Transportation Systems}}
	({{ITSC}})}, \bibinfo{publisher}{{IEEE}}. pp. \bibinfo{pages}{2118--2125}.
\bibitem[{Lee et~al.(2015)Lee, Zhong, Kim, Dimitrijevic, Du and
	Gutesa}]{lee2015}
\bibinfo{author}{Lee, J.}, \bibinfo{author}{Zhong, Z.}, \bibinfo{author}{Kim,
	K.}, \bibinfo{author}{Dimitrijevic, B.}, \bibinfo{author}{Du, B.},
\bibinfo{author}{Gutesa, S.}, \bibinfo{year}{2015}.
\newblock \bibinfo{title}{Examining the applicability of small quadcopter drone
	for traffic surveillance and roadway incident monitoring}, in:
\bibinfo{booktitle}{Transportation {{Research Board}} 94th {{Annual
			Meeting}}}.
\bibitem[{Lu and Elefteriadou(2013)}]{lu2013}
\bibinfo{author}{Lu, C.}, \bibinfo{author}{Elefteriadou, L.},
\bibinfo{year}{2013}.
\newblock \bibinfo{title}{An investigation of freeway capacity before and
	during incidents}.
\newblock \bibinfo{journal}{Transportation Letters} \bibinfo{volume}{5},
\bibinfo{pages}{144--153}.
\bibitem[{Pan et~al.(2013)Pan, Demiryurek, Shahabi and Gupta}]{pan2013}
\bibinfo{author}{Pan, B.}, \bibinfo{author}{Demiryurek, U.},
\bibinfo{author}{Shahabi, C.}, \bibinfo{author}{Gupta, C.},
\bibinfo{year}{2013}.
\newblock \bibinfo{title}{Forecasting spatiotemporal impact of traffic
	incidents on road networks}, in: \bibinfo{booktitle}{2013 {{IEEE}} 13th
	{{International Conference}} on {{Data Mining}}},
\bibinfo{publisher}{{IEEE}}. pp. \bibinfo{pages}{587--596}.
\bibitem[{Parkany and Xie(2005)}]{parkany2005}
\bibinfo{author}{Parkany, E.}, \bibinfo{author}{Xie, C.}, \bibinfo{year}{2005}.
\newblock \bibinfo{title}{A Review of Incident Detection Technologies,
	Algorithms and Their Deployments: What Works and What Doesn't}.
\newblock \bibinfo{type}{Technical Report} \bibinfo{number}{NETCR37}.
{University of Massachusetts Transportation Center}.
\bibitem[{Payne et~al.(1976)Payne, Helfenbein and Knobel}]{payne1976}
\bibinfo{author}{Payne, H.J.}, \bibinfo{author}{Helfenbein, E.D.},
\bibinfo{author}{Knobel, H.C.}, \bibinfo{year}{1976}.
\newblock \bibinfo{title}{Development and Testing of Incident Detection
	Algorithms, Vol. 2: {{Research}} Methodology and Detailed Results}.
\newblock \bibinfo{type}{Technical Report} \bibinfo{number}{FHWA-RD-76-20}.
{Federal Highway Administration}.
\bibitem[{Quiroga et~al.(2004)Quiroga, Kraus, Pina, Hamad and
	Park}]{quiroga2004}
\bibinfo{author}{Quiroga, C.}, \bibinfo{author}{Kraus, E.},
\bibinfo{author}{Pina, R.}, \bibinfo{author}{Hamad, K.},
\bibinfo{author}{Park, E.S.}, \bibinfo{year}{2004}.
\newblock \bibinfo{title}{Incident Characteristics and Impact on Freeway
	Traffic}.
\newblock \bibinfo{type}{Technical {{Report}}} \bibinfo{number}{0-4745}. {Texas
	Transportation Institute}.
\bibitem[{Sim and Roy(2005)}]{sim2005}
\bibinfo{author}{Sim, R.}, \bibinfo{author}{Roy, N.}, \bibinfo{year}{2005}.
\newblock \bibinfo{title}{Global {{A}}-optimal robot exploration in {{SLAM}}},
in: \bibinfo{booktitle}{Proceedings of the 2005 {{IEEE International
			Conference}} on {{Robotics}} and {{Automation}}},
\bibinfo{publisher}{{IEEE}}. pp. \bibinfo{pages}{661--666}.
\bibitem[{Skabardonis et~al.(2003)Skabardonis, Varaiya and
	Petty}]{skabardonis2003}
\bibinfo{author}{Skabardonis, A.}, \bibinfo{author}{Varaiya, P.},
\bibinfo{author}{Petty, K.}, \bibinfo{year}{2003}.
\newblock \bibinfo{title}{Measuring recurrent and nonrecurrent traffic
	congestion}.
\newblock \bibinfo{journal}{Transportation Research Record: Journal of the
	Transportation Research Board} \bibinfo{volume}{1856},
\bibinfo{pages}{118--124}.
\bibitem[{Stephanedes and Chassiakos(1993)}]{stephanedes1993}
\bibinfo{author}{Stephanedes, Y.J.}, \bibinfo{author}{Chassiakos, A.P.},
\bibinfo{year}{1993}.
\newblock \bibinfo{title}{Freeway incident detection through filtering}.
\newblock \bibinfo{journal}{Transportation Research Part C: Emerging
	Technologies} \bibinfo{volume}{1}, \bibinfo{pages}{219--233}.
\bibitem[{Stephanedes and Liu(1995)}]{stephanedes1995}
\bibinfo{author}{Stephanedes, Y.J.}, \bibinfo{author}{Liu, X.},
\bibinfo{year}{1995}.
\newblock \bibinfo{title}{Artificial neural networks for freeway incident
	detection}.
\newblock \bibinfo{journal}{Transportation Research Record} ,
\bibinfo{pages}{91--97}.
\bibitem[{Stevens and Blackstock(2017)}]{stevens2017}
\bibinfo{author}{Stevens, C.R.}, \bibinfo{author}{Blackstock, T.},
\bibinfo{year}{2017}.
\newblock \bibinfo{title}{Demonstration of Unmanned Aircraft Systems Use for
	Traffic Incident Management}.
\newblock \bibinfo{type}{Technical Report}. {Texas A\&M Transportation
	Institute}.
\bibitem[{Sun et~al.(2017)Sun, Jin and Ritchie}]{sun2017}
\bibinfo{author}{Sun, Z.}, \bibinfo{author}{Jin, W.}, \bibinfo{author}{Ritchie,
	S.G.}, \bibinfo{year}{2017}.
\newblock \bibinfo{title}{Simultaneous estimation of states and parameters in
	{{Newell}}'s simplified kinematic wave model with {{Eulerian}} and
	{{Lagrangian}} traffic data}.
\newblock \bibinfo{journal}{Transportation Research Part B: Methodological}
\bibinfo{volume}{104}, \bibinfo{pages}{106--122}.
\bibitem[{Sun et~al.(2019)Sun, Lin, Li and Xiang}]{sun2019}
\bibinfo{author}{Sun, L.}, \bibinfo{author}{Lin, Z.}, \bibinfo{author}{Li, W.},
\bibinfo{author}{Xiang, Y.}, \bibinfo{year}{2019}.
\newblock \bibinfo{title}{Freeway incident detection based on set theory and
	short-range communication}.
\newblock \bibinfo{journal}{Transportation Letters} \bibinfo{volume}{11},
\bibinfo{pages}{558--569}.
\bibitem[{Tamp{\`e}re and Immers(2007)}]{tampere2007}
\bibinfo{author}{Tamp{\`e}re, C.M.}, \bibinfo{author}{Immers, L.H.},
\bibinfo{year}{2007}.
\newblock \bibinfo{title}{An extended kalman filter application for traffic
	state estimation using {{CTM}} with implicit mode switching and dynamic
	parameters}, in: \bibinfo{booktitle}{2007 {{Intelligent Transportation
			Systems Conference}}}, \bibinfo{publisher}{{IEEE}}. pp.
\bibinfo{pages}{209--216}.
\bibitem[{Teng and Qi(2003)}]{teng2003}
\bibinfo{author}{Teng, H.}, \bibinfo{author}{Qi, Y.}, \bibinfo{year}{2003}.
\newblock \bibinfo{title}{Application of wavelet technique to freeway incident
	detection}.
\newblock \bibinfo{journal}{Transportation Research Part C: Emerging
	Technologies} \bibinfo{volume}{11}, \bibinfo{pages}{289--308}.
\bibitem[{Ucinski(2004)}]{ucinski2004}
\bibinfo{author}{Ucinski, D.}, \bibinfo{year}{2004}.
\newblock \bibinfo{title}{Optimal Measurement Methods for Distributed Parameter
	System Identification}.
\newblock \bibinfo{publisher}{{CRC Press}}.
\bibitem[{Wang et~al.(2016a)Wang, Fan and Work}]{wang2016a}
\bibinfo{author}{Wang, R.}, \bibinfo{author}{Fan, S.}, \bibinfo{author}{Work,
	D.B.}, \bibinfo{year}{2016}a.
\newblock \bibinfo{title}{Efficient multiple model particle filtering for joint
	traffic state estimation and incident detection}.
\newblock \bibinfo{journal}{Transportation Research Part C: Emerging
	Technologies} \bibinfo{volume}{71}, \bibinfo{pages}{521--537}.
\bibitem[{Wang and Work(2014)}]{wang2014}
\bibinfo{author}{Wang, R.}, \bibinfo{author}{Work, D.B.}, \bibinfo{year}{2014}.
\newblock \bibinfo{title}{Interactive multiple model ensemble {{Kalman}} filter
	for traffic estimation and incident detection}, in: \bibinfo{booktitle}{17th
	{{International IEEE Conference}} on {{Intelligent Transportation Systems}}
	({{ITSC}})}, \bibinfo{publisher}{{IEEE}}. pp. \bibinfo{pages}{804--809}.
\bibitem[{Wang et~al.(2016b)Wang, Work and Sowers}]{wang2016}
\bibinfo{author}{Wang, R.}, \bibinfo{author}{Work, D.B.},
\bibinfo{author}{Sowers, R.}, \bibinfo{year}{2016}b.
\newblock \bibinfo{title}{Multiple model particle filter for traffic estimation
	and incident detection}.
\newblock \bibinfo{journal}{IEEE Transactions on Intelligent Transportation
	Systems} \bibinfo{volume}{17}, \bibinfo{pages}{3461--3470}.
\bibitem[{Wang and Papageorgiou(2005)}]{wang2005b}
\bibinfo{author}{Wang, Y.}, \bibinfo{author}{Papageorgiou, M.},
\bibinfo{year}{2005}.
\newblock \bibinfo{title}{Real-time freeway traffic state estimation based on
	extended {{Kalman}} filter: {{A}} general approach}.
\newblock \bibinfo{journal}{Transportation Research Part B: Methodological}
\bibinfo{volume}{39}, \bibinfo{pages}{141--167}.
\bibitem[{Willsky et~al.(1980)Willsky, Chow, Gershwin, Greene, Houpt and
	Kurkjian}]{willsky1980}
\bibinfo{author}{Willsky, A.}, \bibinfo{author}{Chow, E.},
\bibinfo{author}{Gershwin, S.}, \bibinfo{author}{Greene, C.},
\bibinfo{author}{Houpt, P.}, \bibinfo{author}{Kurkjian, A.},
\bibinfo{year}{1980}.
\newblock \bibinfo{title}{Dynamic model-based techniques for the detection of
	incidents on freeways}.
\newblock \bibinfo{journal}{IEEE Transactions on Automatic Control}
\bibinfo{volume}{25}, \bibinfo{pages}{347--360}.

\end{thebibliography}

}
\end{document}